%% file: final_file.tex
\documentclass[10pt,journal,compsoc]{IEEEtran}

\usepackage{cite}
\usepackage{graphicx}
\usepackage[fleqn]{amsmath}
\usepackage{algorithmic}
\usepackage{url}

\usepackage[numbers,sort&compress]{natbib}
\usepackage{color,soul}
\usepackage{textcomp}
\usepackage{xcolor}
\usepackage{hyperref}
\usepackage{xspace}
\usepackage{wasysym}
\usepackage{arydshln}
\usepackage{subfig}
\usepackage{amsthm}
\usepackage{amssymb,amsfonts}
\usepackage{stfloats}
\usepackage{nccmath}
\usepackage[ruled,norelsize]{algorithm2e}

\usepackage{pgfplots}
\pgfplotsset{compat=1.8}
\usepgfplotslibrary{statistics}
\usepackage{tikz}
\usetikzlibrary{shapes}
\usetikzlibrary{matrix}
\usetikzlibrary{patterns}
\usetikzlibrary{backgrounds}
\usetikzlibrary{scopes}
\usepackage{multirow}

\makeatletter
\newcommand{\removelatexerror}{\let\@latex@error\@gobble}
\makeatother

\newcommand{\figref}[1]{\mbox{Fig.~\ref{#1}}}
\newcommand{\tabref}[1]{\mbox{TABLE~\ref{#1}}}

\newcommand{\neck}[1]{{\vspace{3pt}\noindent{\textbf{#1}}}}

\begin{document}
\title{Reproducible and Portable Big Data Analytics\\ in the Cloud}

\author{Xin~Wang,~Pei~Guo,~Xingyan~Li, \\Aryya~Gangopadhyay,~Carl~E.~Busart,~Jade~Freeman,~and~Jianwu~Wang$^*$
\IEEEcompsocitemizethanks{\IEEEcompsocthanksitem $^*$Corresponding author\\
\IEEEcompsocthanksitem X. Wang, P. Guo, X. Li, A. Gangopadhyay and J. Wang are with the Department of Information Systems, University of Maryland, Baltimore County, Baltimore, MD, 21250.\\E-mail: \{xinwang11, peiguo1, xingyanli, gangopad, jianwu\}@umbc.edu.\protect\\
\IEEEcompsocthanksitem C. Busart and J. Freeman are with the DEVCOM Army Research Laboratory, Adelphi, MD, 20783.\\E-mail: \{carl.e.busart.civ, jade.l.freeman2.civ\}@army.mil.}
}

\IEEEtitleabstractindextext{%
\begin{abstract}
Cloud computing has become a major approach to help reproduce computational experiments. Yet there are still two main difficulties in reproducing batch based big data analytics (including descriptive and predictive analytics) in the cloud. The first is how to automate end-to-end scalable execution of analytics including distributed environment provisioning, analytics pipeline description, parallel execution, and resource termination. The second is that an application developed for one cloud is difficult to be reproduced in another cloud, a.k.a. vendor lock-in problem. To tackle these problems, we leverage serverless computing and containerization techniques for automated scalable execution and reproducibility, and utilize the adapter design pattern to enable application portability and reproducibility across different clouds. We propose and develop an open-source toolkit that supports 1) fully automated end-to-end execution and reproduction via a single command, 2) automated data and configuration storage for each execution, 3) flexible client modes based on user preferences, 4) execution history query, and 5) simple reproduction of existing executions in the same environment or a different environment. We did extensive experiments on both AWS and Azure using four big data analytics applications that run on virtual CPU/GPU clusters. The experiments show our toolkit can achieve good execution performance, scalability, and efficient reproducibility for cloud-based big data analytics.

\end{abstract}

}

\maketitle

%
\IEEEdisplaynontitleabstractindextext

\IEEEpeerreviewmaketitle

\IEEEraisesectionheading{\section{Introduction}}
\IEEEPARstart{R}eproducibility is increasingly required by the research community, funding agencies, and publishers~\cite{NAP25303}. By reproducing an existing computational experiment and obtaining consistent results, we can have more confidence in the research. Further, besides reproducing the exact process, it is also valuable to explore how the experiment behaves with different input datasets, execution arguments, and environments. Cloud computing has been a major approach for reproducibility~\cite{qasha2016framework} because cloud services can be leveraged to provision data, software, or hardware needed in reproduction. 
For instance, paper~\cite{Bresearch} summarized 13 aspects that cloud computing can help with reproducibility. 

In this paper, we mainly address the following challenges in cloud-based reproducibility. First, it is still difficult to achieve end-to-end automated big data analytics execution and reproduction in the cloud. The end-to-end automation should support scale-up and scale-out of distributed hardware environment, software environment provisioning, data and configuration storage for each execution, resource termination after execution, execution history query and reproducibility of existing executions in the same environment or a different cloud environment. Second, because the services provided by each service provider such as AWS and Azure are proprietary, an application developed for one cloud cannot run in another cloud, which is a well-known vendor lock-in challenge. Two scientific problems to be studied by tackling challenges are: 1) what is a proper abstraction and design for better reproducibility support from both user and toolkit perspectives, 2) what is a more efficient way to achieve cloud-based reproducibility for big data analytics. We note our work only supports batch based processing big data analytics jobs, including descriptive and predictive analytics, not interactive jobs like database queries.

Based on the above challenges and scientific problems, we propose an approach and corresponding open-source toolkit~\cite{Reproducible_and_Portable_Big_Data_Analytics_in_Cloud} for \underline{R}eproducible and \underline{P}ortable big data \underline{A}nalytics in the \underline{C}loud (RPAC).
Our contributions are summarized as follows.

\begin{itemize}
    \item Our proposed approach and toolkit integrate serverless computing techniques to automate end-to-end batch based big data analytics execution. Tasks of big data analytics execution (resource provisioning, application execution, data storage and resource termination) are encapsulated as cloud functions and automatically triggered by proper events. With the full automation support, users can re-run the exact execution or run the application with different configurations, including different scale-out and scale-up factors, via only one command. Our RPAC toolkit supports both AWS and Azure cloud environments.
    \item For easy reproducibility, we make proper data modeling and abstraction. It first separates essential information required for reproducibility and detailed information required by each cloud provider. Following the separation of concerns principle, it further separates the essential information into three categories (resources, application, personal) for easy reconfiguration. The essential information will also be automatically stored in the cloud by our toolkit as authentic recording of the execution. Later, the storage URL can be published and shared as the single source to reproduce the historical execution.
    \item To deal with the vendor lock-in challenge, on top of the above abstractions, we propose a \underline{C}loud \underline{A}gnostic \underline{A}pplication \underline{M}odel~(CAAM) to support execution and reproducibility portability with different cloud providers. CAAM abstracts the application out of its cloud specific logic, and allows reproducing executions in another cloud via only minimal configuration changes from the user.
    \item We benchmark both CPU-based and GPU-based big data analytics applications using our RPAC toolkit. We measure the overhead of data storing for reproducibility. We also did extensive experiments to benchmark three applications on different cloud providers in terms of execution performance, scalability and reproducibility efficiency.
\end{itemize}

The rest of the paper is organized as follows. In Section \ref{sec:background}, we briefly introduce related techniques our work is built on. Section \ref{sec:architecture} provides an overview of our proposed approach. Three main parts of our approach, namely data modeling, automated execution and reproduction of big data analytics in the cloud are explained in Section \ref{datamodeling}, Section \ref{sec:automation} and Section \ref{sec:reproduction}, respectively. Experiments and benchmarking results are discussed in Section \ref{sec:experiments}. We compare our work with related studies in \ref{sec:relatedworks} and conclude in Section \ref{sec:conclusions}.

\section{Background}
\label{sec:background}

\subsection{Big Data Analytics} 
\label{sec:bigdata_background}
To deal with increasing data volumes in data analytics, many platforms have been proposed to achieve parallelization of the analytics in a distributed environment. We explain three such platforms that our work is built on for reproducibility. As one of the most popular big data platform, Spark~\cite{spark-url} follows and extends the MapReduce paradigm~\cite{dean2008mapreduce} and achieve parallelism by distributing input data among many parallel tasks of the same function. To run an application, Spark employs a master process on one node and a worker process on each of other nodes so the worker processes can take tasks from the master process and run them in parallel.
Similar to Spark, a Dask~\cite{dask-url} application is composed as a task graph that can be distributed within one computer or a distributed computing environment. Dask employs a similar master-worker framework for task scheduling. 
Horovod~\cite{horovod-url}, as a popular software framework for distributed learning, provides data parallel deep learning optimized for GPU-based data analytics. For coordinating execution between distributed processes on GPU, Horovod can use Message Passing Interface (MPI) for communicating data with high performance. The CUDA-aware MPI is commonly used in HPC to build applications that can scale to multi-node computer clusters~\cite{kraus2013introduction}.

\subsection{Reproducibility} 
\label{sec:reproduce_background}
There have been many definitions of reproducibility and similar terms like replicability and repeatability~\cite{barba2018terminologies,acm-reproducibility}. Unfortunately, these definitions are not very consistent, some even contradict with each other~\cite{NAP25303}. Here, we simply define reproducibility as a capability that obtains consistent results using the same computational steps, methods, and code. As paper~\cite{bartusch2019reproducible} said, containerization is one of the valid and common solutions for the reproducible software deployment problem of scientific pipelines. 
For cloud-based reproducibility, it studies how to re-execute an existing application in the cloud~\cite{Bresearch}. We categorize reproducibility support into four ways: 1) rerun exactly the same application with the same hardware and software environment, 2) reproduce with a different application configuration to know how the application performs with different datasets or arguments, 3) reproduce with different cloud provider hardware environment (virtual machine type and number, etc.) within the same cloud provider to test scale-up and scale-out; and 4) reproduce with a different cloud provider to avoid vendor lock-in problem. Our toolkit is built to support all four types of reproduction.

\subsection{Serverless Computing}
\label{sec:serverless_background}

As a recent cloud-based execution model, serverless computing provides a few advantages. First, it responds to user service requests without maintaining back-end servers in the cloud. Second, it employs Function as a Service (FaaS) architecture that allows customers to develop separate functions directly rather than standalone cloud applications. As explained in~\cite{mcgrath2017serverless}, each application logic/pipeline is split into functions and application execution is based on internal or external events. All major cloud providers offer serverless services, including AWS Lambda, Azure Functions and Google Cloud Functions.

\section{Overview of Reproducible and Portable Data Analytics in the Cloud} 
\label{sec:architecture}

In this section, we provide an overview of how our proposed approach achieves reproducible and portable data analytics in the cloud. With the approach and corresponding open-source toolkit RPAC for reproducible and portable data analytics in the cloud, users can easily re-run previous experiments with the same or different setups including environments, application arguments, input data and cloud providers. Our approach is built on top of serverless computing and we adopt a new way of utilizing serverless computing for large scale computations. So we will explain first how to use serverless large scale computations, then how to use serverless for big data analytics reproducibility.

\subsection{Serverless based Reproducibility}
\label{sec:reproducibilityoverview}
\begin{figure*}[t]
    \centering
    \vspace{-6pt}
    \includegraphics[width=0.7\textwidth]{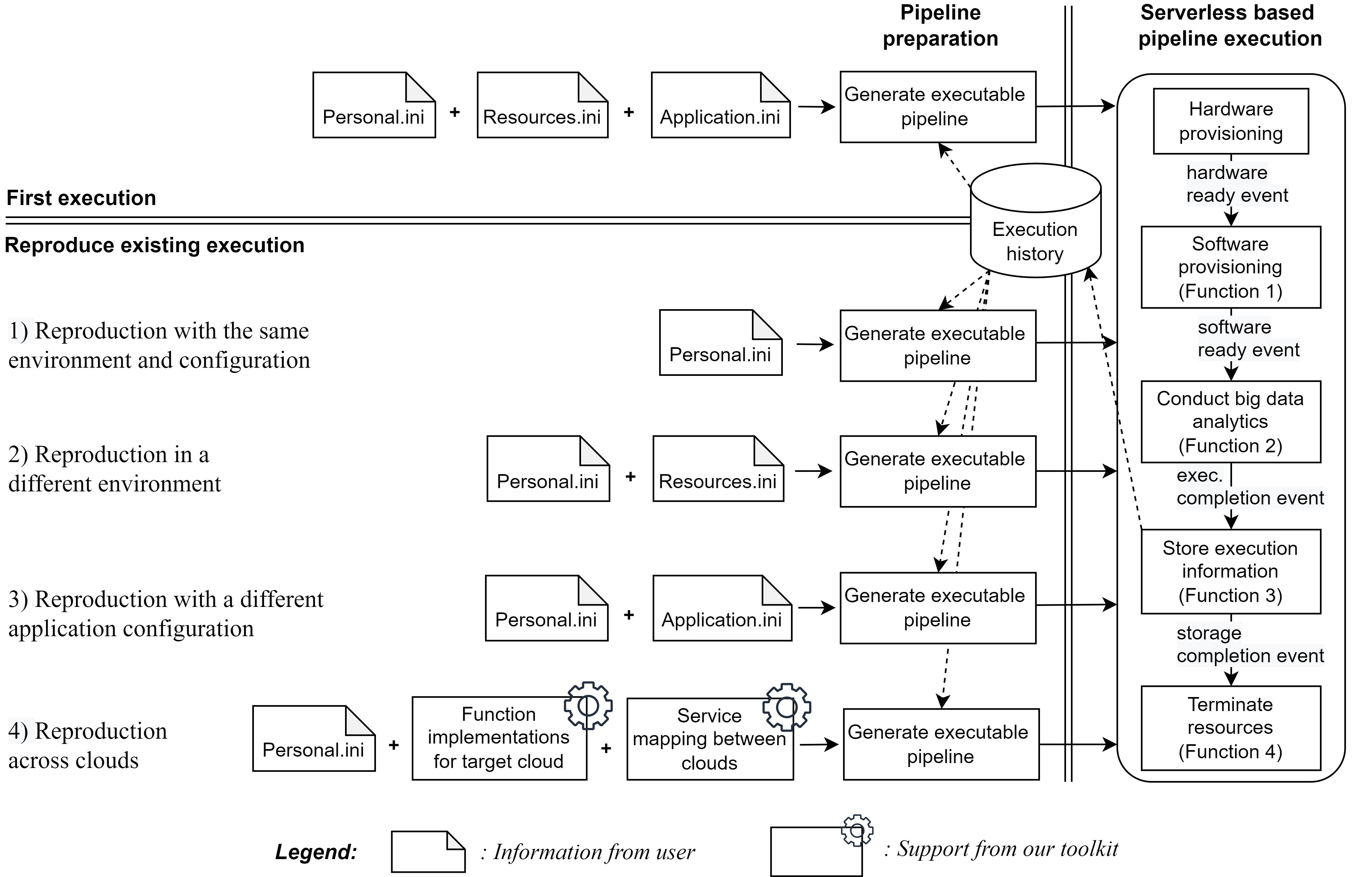}
    \vspace{-6pt}
    \caption{The overview of our proposed approach for reproducible and portable data analytics in the cloud.}
    \label{fig:life-cycle}
    \vspace{-9pt}
\end{figure*}

As shown in \figref{fig:life-cycle}, our proposed approach has two parts: 1) first execution of an application, and 2) reproduction of the existing execution from historical configurations. Both first execution and reproduction are automated via serverless-based approach shown on the right. 

In the beginning, there is no execution history for querying and reproducing. Clients need to prepare configurations to generate the pipeline file for the whole execution. The configuration includes all configurable setups, for example, the application-based information like application programs, arguments, input data, and the cloud-based information like virtual cluster type, size, network setting, memory, with personal credentials. Our toolkit will take this information to create an executable pipeline for a target cloud. With this pipeline, the data analytics application will execute in the cloud environment, output its results to the storage, and automatically terminate resources once the execution finishes. We will explain in detail how we leverage serverless techniques for automated big data analytics in Section~\ref{sec:automation}.  

After an application is executed, clients can reproduce it based on its execution history. Our RPAC toolkit will generate a pipeline file based on the execution history and reproduction configurations. If the client wants to reproduce an existing execution with the exact environment and configuration, the pipeline file within the execution history can be used directly by our toolkit for reproducibility. If the client chooses to reproduce existing execution within the same cloud, but with a different environment or application, our toolkit will combine changed configurations of cloud resources or applications with the historical execution information to generate a new pipeline file. If the client prefers reproducing existing execution on a different cloud, our toolkit will provide cloud service mapping and implementations of functions in the target cloud. With user-provided personal information and historical execution, a new pipeline will be generated for the target cloud. Finally, with the pipeline file executable by cloud serverless services, the data analytics will be reproduced in cloud automatically. Details of how our approach achieves reproducibility will be explained in Section~\ref{sec:reproduction}.

We would like to note the serverless pipeline used here is different from most other workflow or pipeline definitions such as~{\cite{bartusch2019reproducible}\cite{baldini2016cloud}\cite{babuji2019parsl}\cite{stubbs2021tapis}}. These definitions only include the processing steps and their dependencies. They do not describe how to provision hardware and software environments because they assume these environments are ready before pipeline execution. Our serverless pipeline includes the full execution life cycle including hardware and software provisioning, big data analytics, execution export and resource release. Our pipeline does not describe internal processing steps, but could be integrated with traditional pipelines as internal logic description in its Function 2: conduct big data analytics.

\subsection{Serverless based Large Scale Application in the Cloud}
\label{sec:serverless4reproducibility}

Traditionally, serverless computing is used to execute serverless pipelines and the functions defined in each pipeline directly via cloud services like AWS CloudFormation. In this case, the computation is executed following the pipeline without using any additional cloud resources. Because of the memory and CPU limit for serverless functions, this approach can only handle computations whose resource requirements are light. For instance, OpenWhisk~{\cite{openwhisk}} is an event-based serverless computing cloud platform, which allows users to implement their own OpenWhisk APIs for the connections between the event source and trigger, the trigger rule, and the computation actions.

\begin{figure}[t]
    \centering
    \vspace{-3pt}
    \includegraphics[width=0.47\textwidth]{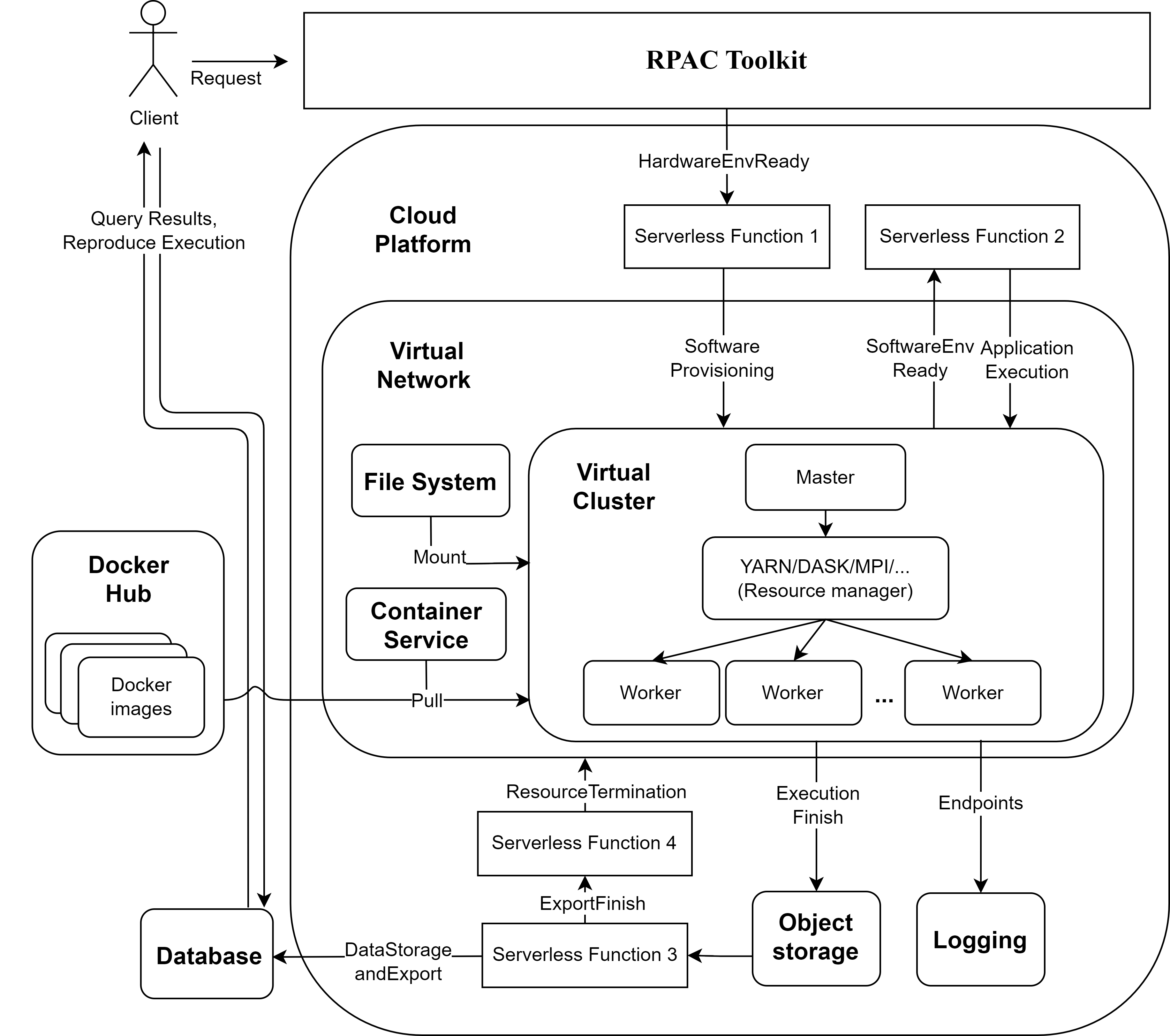}
    \vspace{-3pt}
    \caption{The interaction between our proposed toolkit and cloud resources.}
    \label{fig:overview}
    \vspace{-3pt}
\end{figure}

Different from the above way to use serverless, we leverage serverless computing and its FaaS to achieve reproducibility for big data analytics in the cloud. The main difference is that we use the serverless pipeline as a way to orchestrate and manage additional cloud resources for heavy workloads while each step is wrapped as a function. In this way, both the serverless pipeline and its functions do not execute heavy commands directly. Instead, each function's execution only submits commands from serverless to the additional cloud resources. 
Then when the function is triggered, the commands will be transferred to the additional cloud services and be executed as background processes so they can return without waiting for the finish of the commands.
As shown in {\figref{fig:overview}}, the serverless pipeline listens to events sent by our toolkit or other cloud services. By mapping the triggers mentioned in the event with the trigger rule associated with each serverless function, it knows which serverless function will be involved based on the received event. For instance, Function 2 will be triggered when the pipeline receives an event as SoftwareEnvReady. For each function's execution, it only submits commands from serverless to additional cloud services such as AWS EC2, so the resource and time limits for serverless functions will not be violated. Also, all major cloud providers including AWS and Azure, only enforce time limits for serverless functions, not serverless pipelines. So serverless pipelines are capable of large-scale computations that might take a long time. 
Serverless pipeline is also a reasonable choice from a budgetary cost perspective because serverless service is charged by the number of function invocations and the duration it takes to execute, not the deployment time of the pipeline.

\section{Data Modeling and Storage for Reproducibility}
\label{datamodeling}

\begin{figure*}[b]
    \centering
    \vspace{-3pt}
    \includegraphics[width=0.74\textwidth]{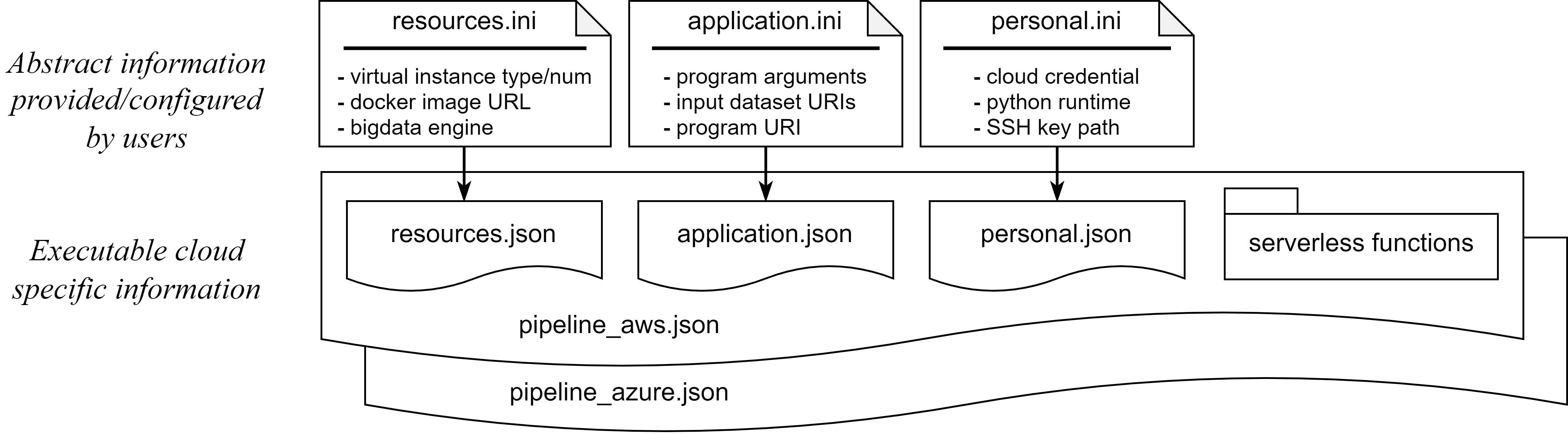}
    \vspace{-6pt}
    \caption{Data modeling and abstraction for reproducibility.}
    \label{fig:datamodeling}
    \vspace{-6pt}
\end{figure*}

\begin{figure*}[ht] \small
	\centering
	\fbox{\parbox{6.5in}{\vspace{0ex}
	\parbox[b]{6.5in}{
		$<$abstract$\_$req$\_$4$\_$big$\_$data$\_$analytics$>$ ::= $<resources>,<personal>,<application>$ \\
		$<resources>$ ::= $<bigdata\_engine>,<cloud.aws>\mid<cloud.azure>,<reproduce>$ \\
		$<bigdata\_engine> ::=$ "$none$"$\mid$"$spark$"$\mid$"$horovod$"$\mid$"$dask$"	\\
		$<cloud.aws> ::= <region>,<instance\_number>,<subnet\_id>,< instance\_type>,<vpc\_id>$ \\
		$<cloud.azure> ::= <region>,<instance\_number>,<resource\_group\_name>,<instance\_type>$ \\
		$<reproduce> ::= <reproduce\_storage>,<reproduce\_database>$ \\
		$<personal> ::= <cloud\_provider>,<key\_path>,<key\_name>,<python\_runtime>,<cloud\_credentials>$ \\
		$<application> ::= <docker\_image>,<data\_uri>,<command>,<bootstrap>$
	}}}
	\vspace{-0.5em}
	\caption{Normative form for abstract request information.}
	\vspace{-3pt}
	\label{fig:form}
\end{figure*}

\begin{figure*}[hb]
    \centering
    \includegraphics[width=0.98\textwidth]{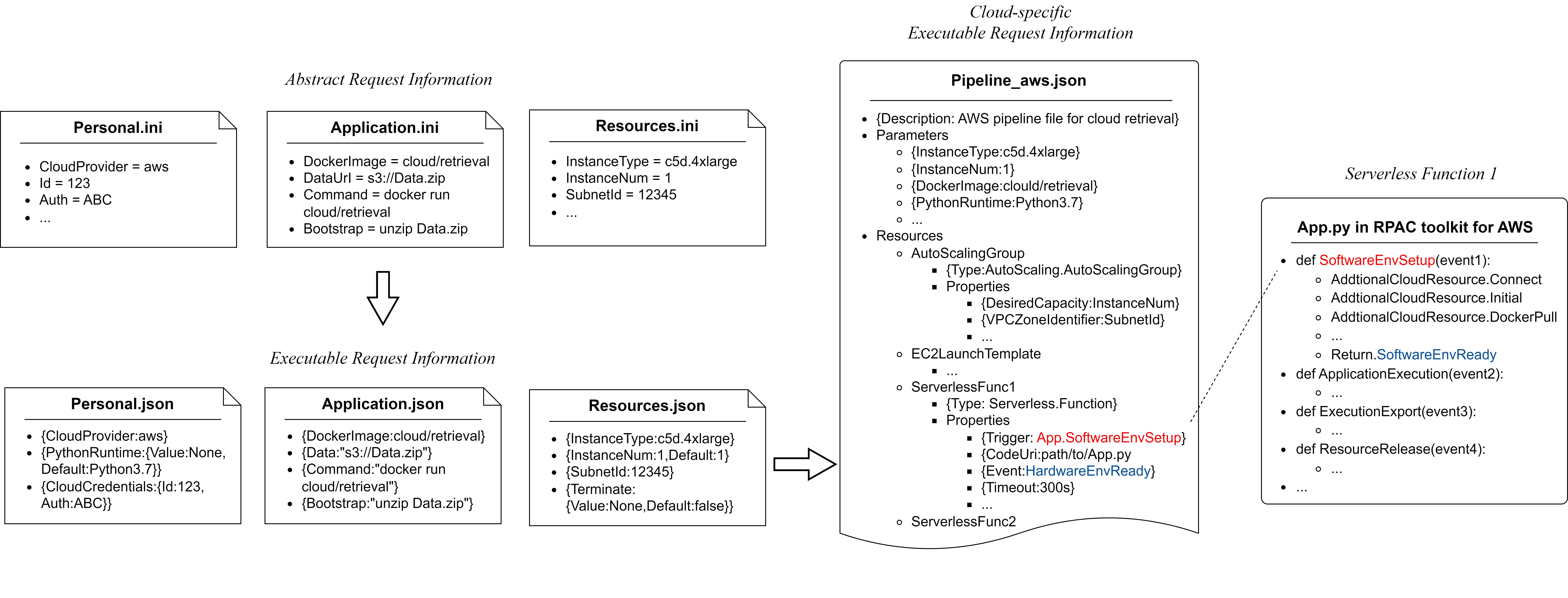}
    \vspace{-9pt}
    \caption{Mapping from abstract requirement to executable cloud specific serverless pipeline and functions.}
    \label{fig:PipelineFile}
\end{figure*}

To achieve easy configurability by users and future reproducibility across cloud providers, we categorize data based on their usage and employ different levels of data abstraction. Specifically, the data model contains three parts: abstract request information, executable request information and execution history information. We believe the data model can serve as a reference model for different reproducibility toolkits.  

\subsection{Abstract Request Information}
\label{sec:abstracinfo}
To avoid learning specific specifications and templates for specific clouds, we extract minimal information a user has to provide for application execution or reproduction. Further, as shown in the upper part of \figref{fig:datamodeling}, we categorize the information into three separate key-value based configuration files where \texttt{\small ini} is used as file extension to distinguish them from other file types used by our toolkit. Specifically, \texttt{\small resources.ini} stores hardware and software resources information such as virtual machine type, virtual instance number, docker image URL and big data engine; \texttt{\small application.ini} records the program URI of the application, program arguments, and input dataset URIs of the program; \texttt{\small personal.ini} contains the cloud credential information such as SSH key location and cloud credential info (which can also be provided at runtime for security concerns). We use three different files so only a subset of files needs to be edited for each type of reproducibility shown in \figref{fig:life-cycle}. A complete and formal listing of the information can be found at {\figref{fig:form}} using syntax of Backus-naur form (BNF)~{\cite{mccracken2003backus}}.   

\subsection{Executable Request Information} We separate information that is required for actual cloud-based application execution into four files and use \texttt{\small json} as the file extension. Such files have to follow specifications set by each cloud. For such files, our RPAC toolkit generates them automatically based on corresponding abstract \texttt{\small ini} file(s) mentioned above.
The first file is \texttt{\small resources.json} which describes hardware and software environment info. This file has to be changed if the cloud provider is switched. The \texttt{\small resources.json} will be generated based on the above \texttt{\small resources.ini} file, the cloud type and the type of big data analytics. Another file is \texttt{\small application.json} which contains application specific information and will be generated by our toolkit based on the above \texttt{\small application.ini} file and the cloud type. Similarly, \texttt{\small personal.json} can be generated from \texttt{\small personal.ini}. As shown in \figref{fig:datamodeling}, by combining \texttt{\small resources.json}, \texttt{\small application.json}, \texttt{\small personal.json} and four cloud-specific serverless functions shown in \figref{fig:life-cycle}, we get \texttt{\small pipeline.json} that describes the execution logic of the serverless application. Our RPAC toolkit contains template \texttt{\small json} files and serverless function implementations so they can be reused for different data analytics applications. 
We illustrate how to map from abstract requirements in Section~{\ref{sec:abstracinfo}} to an executable cloud specific serverless pipeline and its implemented functions in {\figref{fig:PipelineFile}}. The pipeline file is generated by the three abstract request information provided by users, which transfers the stateless configurations to executable cloud-specific information. The abstract request information is first transferred to the executable request information while missing parameters can be filled with their default values. All parameters in the executable request information are also sorted out based on the cloud-specific schema. By combining with corresponding serverless functions, the cloud-specific executable request information, like {\texttt{\small{pipeline$\_$aws.json}}}, will be generated and executed in our RPAC toolkit. Each serverless function listens to the upcoming events. If a received event (e.g.,\textit{HardwareEnvReady} in {\figref{fig:PipelineFile}}) matches, the associated function (e.g., \textit{SoftwareEnvSetup()} in {\figref{fig:PipelineFile}}) will be triggered. At the end of the function execution, a new event (e.g., \textit{SoftwareEnvReady} in {\figref{fig:PipelineFile}}) will be returned to trigger the downstream functions.

The differences between the two types of request information are summarized below. Abstract request information, as a user-friendly abstraction, contains the minimal information a user has to provide for application execution or reproduction. In comparison, the executable request information describes the execution logic of the serverless application, which is required for actual cloud-based application execution. Our RPAC toolkit will generate the executable request information based on the corresponding abstract information during the execution.

Next, we will explain how the files are used for automated execution in detail in Section~\ref{sec:serverless_automation} and how they are reused or transformed for reproduction in Section~\ref{sec:reproduction}. 

\begin{figure*}[ht]
    \centering
    \vspace{-3pt}
    \includegraphics[width=0.93\textwidth]{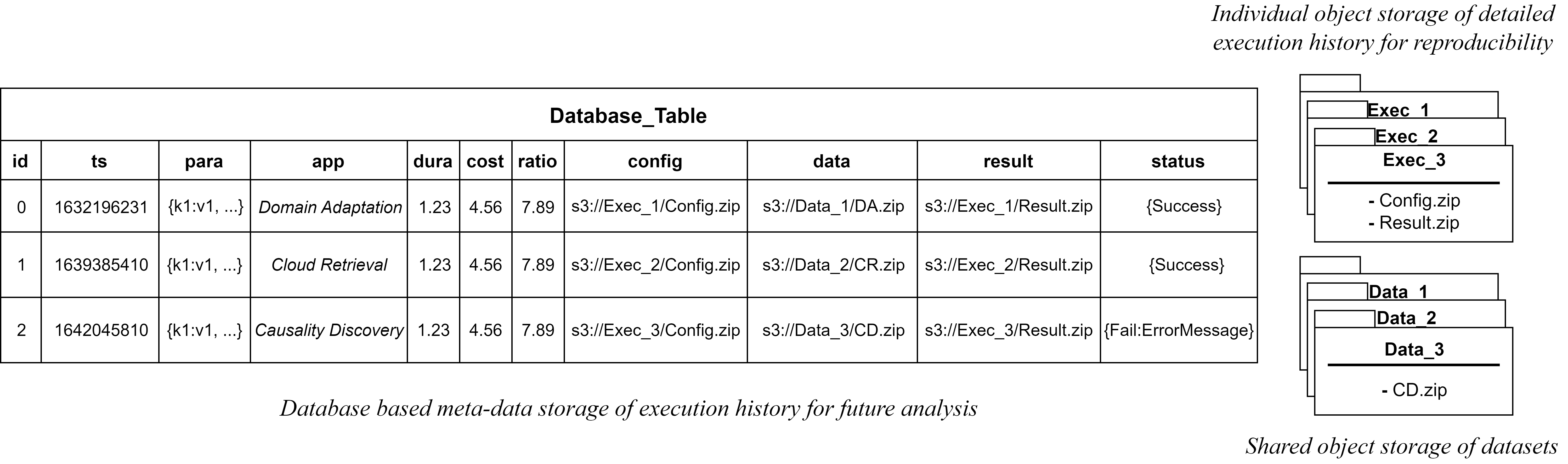}
    \vspace{-6pt}
    \caption{Data modeling of execution history information.}
    \label{fig:dbitem}
    \vspace{-3pt}
\end{figure*}

\subsection{Execution History Information} 

Execution history information is critical to share each execution for later analysis and reproduction. As illustrated in \figref{fig:dbitem}, we classify execution history related data into three categories and store them separately. The first category is execution log metadata, such as timestamps, duration, cost, and status, which are stored in the database for query. Key-value based execution parameters including analytics command line and arguments are also stored for easy comparison among executions. This metadata information is unique for each execution, not required for reproducibility, but useful for later analysis such as finding the fastest execution time of the same application on different clouds or cloud resources. For information that can be referred from external resources, such as input datasets, output files and configuration files used for the execution, only their URLs are stored in the database. The second category is object based storage of each execution information for reproducibility. Two items are stored for each execution: 1) abstract request information (\texttt{\small resources.ini}, \texttt{\small personal.ini} and \texttt{\small application.ini} in Config.zip), 2) execution output datasets in Result.zip. Only abstract request information, not cloud specific information, is stored by our RPAC toolkit in order to minimize storage overhead. These data are compressed, categorized and stored in cloud object storage services such as AWS S3 and Azure Blob storage so a unique URL could be obtained for each execution. Because the data has the complete information to achieve reproducibility, the URL could be easily published as public records following the Research Object framework~\cite{bechhofer2013linked} so it can be referred to via a DOI identifier later as the single source for reproducibility. The third category is shared object storage of input datasets. It is stored separately so that multiple executions with the same input data only need one object storage. Also, cloud storage services like AWS S3 and Azure Blob storage allow automatic versioning so minor changes of input datasets do not require a fully separate storage.

\section{Automated Big Data Analytics in the Cloud towards Reproducibility} 
\label{sec:automation}

To achieve easy reproducibility, the execution should be as automated as possible to minimize manual operations during reproduction phase. Also, an execution should be easily configurable for different scalability factors, application parameters, even cloud providers. In this section, we discuss our techniques to achieve fully automated big data analytics in the cloud so an application can be executed and later reproduced using only one command. 

\subsection{Serverless and Docker-based Execution Automation} 
\label{sec:serverless_automation}

We leverage serverless computing to achieve overall analytics pipeline description and execution, and docker for software environment setup. Serverless computing offers a few advantages for reproducibility: 1) it saves costs because we do not need to maintain a server in the cloud especially for cases reproduction does not happen frequently; 2) its FaaS model allows us to design and implement separate functions required for automated execution/reproduction; 3) its event-based function composition and execution eliminates the requirement of a separate workflow/pipeline software which is needed for many traditional workflow-based reproducibility~\cite{NAP25303}.

\begin{figure*}[htbp]
    \centering
    \vspace{-3pt}
    \includegraphics[width=0.915\textwidth]{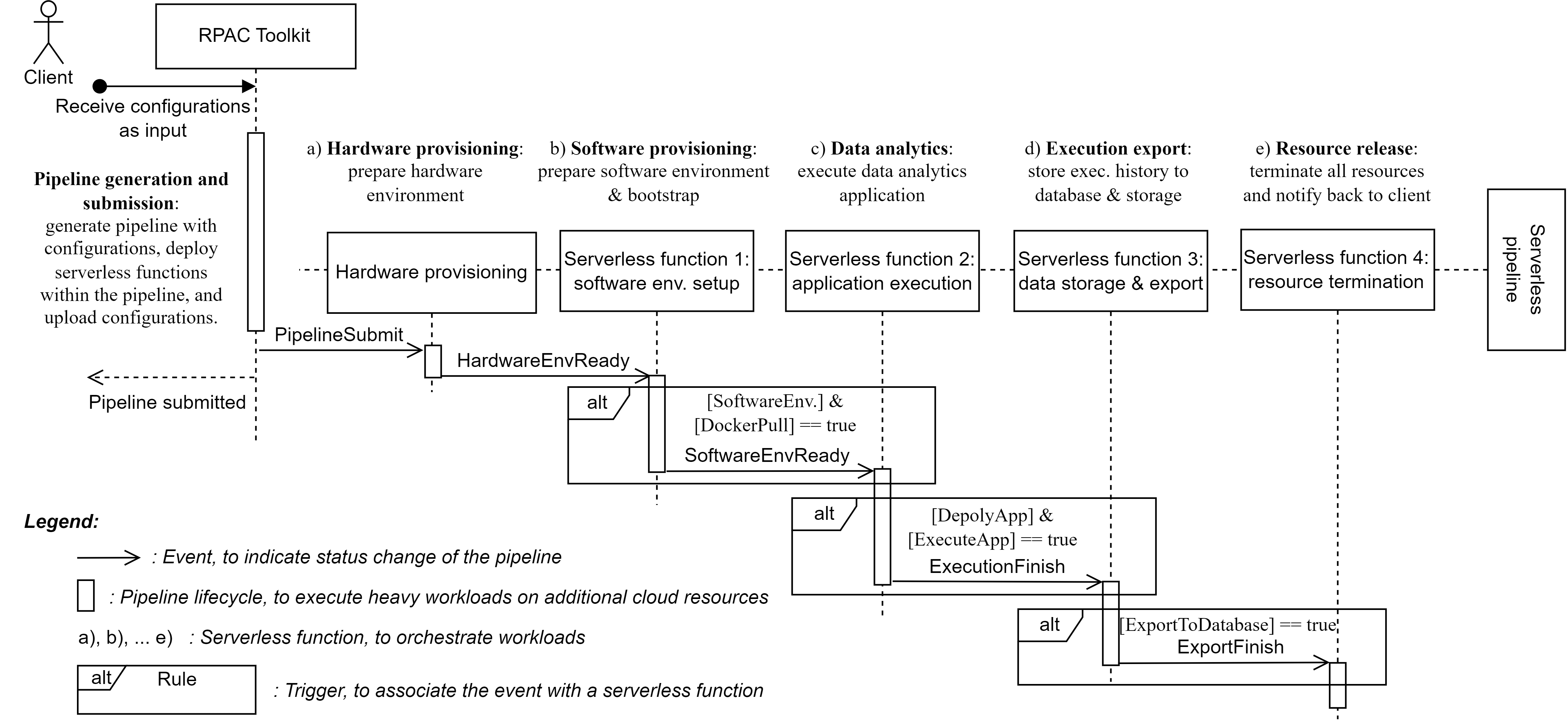}
    \caption{The system sequence diagram for automated execution of big data analytics in serverless framework.}
    \vspace{-3pt}
    \label{fig:timeline}
\end{figure*}

As explained in Section~\ref{sec:serverless_background}, serverless computing offers templates to describe cloud service resources required by the application, structured application pipeline, and event-based execution. Each component in the application pipeline is implemented as a serverless function and triggered by the events it listens to. So the pipeline binds cloud services with the specific event in order to trigger the corresponding serverless function. In addition, we can package complicated software dependencies required for an application via docker. The details of the automation are illustrated in~\figref{fig:timeline}. 

After receiving the user request, RPAC execution automation starts with pipeline generation and submission. Based on configurations, RPAC generates corresponding pipeline files, deploys its serverless functions, and uploads these configurations to the storage except client personal information. RPAC then submits this pipeline to the cloud and starts serverless execution. The serverless pipeline starts with the on-demand hardware environment provisioning (step a in~\figref{fig:timeline}) via cloud manager services (such as CloudFormation for AWS and Deployment Manager for Azure). The hardware provisioning is more like an on-demand resource request service that is a prerequirement for all serverless functions. So we put the hardware provisioning at the beginning of the serverless pipeline.
To conduct big data analytics, we also need to create a virtual cluster by specifying the type and sub-type of virtual machines, the number of virtual machines, network security groups, etc. Cloud manager services allow the information to be submitted based on their semi-structured specification such as JSON and YAML. RPAC will send a reply once the pipeline file is submitted.

The remaining steps of the automated pipeline execution are done via four cloud functions. On top of the virtual hardware environment provisioned, the next automation step is to deploy the required software to run the application (step b in~\figref{fig:timeline}). It is achieved by the first serverless function, which pulls required docker file and starts it.
After the hardware and software environments are provisioned, it is ready to execute applications. The second serverless function in~\figref{fig:timeline} executes the application by deploying user application (e.g., download application codes and unzip them) and running its commands with proper parameters (input data, application specific arguments, etc.). The third function exports all addresses of stored files to cloud database and object storage for future query and reproduction. After the storage completes, a termination event is sent to the last function, which terminates all cloud resources. At this time, the whole pipeline is fully executed, and the client is able to check and query information stored in the database and object storage. 

All these functions are triggered automatically when they receive corresponding events. The cloud manager services mentioned before can help the client manually send events to the serverless function from cloud console. In order to achieve full automation, these events can also be delivered to the target function by cloud event handling services (such as EventBridge for AWS and Event Grid for Azure) using a pre-defined event rule. Each serverless function needs to set up an event rule which specifies what type/property of event can trigger this function. For example, the rule of execution export function (step d in~\figref{fig:timeline}) requires the event source from object storage with a rule-defined prefix, like \textit{export}.

Besides serverless-based execution, our RPAC toolkit also supports cloud SDK-based execution to allow flexible client modes. Their differences are summarized in {\tabref{tab:mode}}. 
The cloud SDK mode is designed based on the cloud-specific software development toolkit (SDK). SDK facilitates the creation of applications by having a compiler, debugger and a software framework based on its functionality. The implementation of this SDK-based mode contains cloud application programming interfaces (APIs) for pipeline management. For example, AWS Boto python SDK can be invoked to describe the status of EC2 using \texttt{\small{ec2.describe$\_$instances()}}. The execution can be automated by a periodical status pulling loop. This SDK-based mode requires programming knowledge and a complete understanding of the data analytics pipeline, so that developers become preferred users since they can either run the application in a fully automated way or step-wise execution for debugging purposes. By supporting different execution modes, users can make flexible choices. In comparison, serverless based approach is fully automated and more efficient because only the execution is managed via internal event triggering. No communications between client and cloud are needed once the pipeline is submitted. 

\begin{table*}[ht]
    \vspace{-3pt}
    \centering
    \caption{Comparison of two execution modes of RPAC toolkit for execution and reproducibility.}
    \vspace{-3pt}
    \resizebox{.985\textwidth}{!}{
    \begin{tabular}{|c|c|c|c|}
    \hline
        \multirow{2}*{\textbf{Execution mode}} & \multirow{2}*{\textbf{Techniques}} & \multirow{2}*{\textbf{Automation}} & \textbf{Preferred} \\
        & & & \textbf{users} \\
        \hline
        \hline
         \multirow{2}*{Cloud SDK}  & Use the local machine terminal for data analytics. & Full automation by & \multirow{2}*{Developer} \\
         & Implementation for the whole execution with cloud SDK. & periodical status pulling from client&\\
         \hdashline[1pt/1pt]
         \multirow{2}*{Serverless} & Use the cloud specific serverless service for data analytics. & Full automation by & End user~/ \\
         & Managed by binding trigger with serverless function. & serverless event triggering within cloud & Developer \\
         \hline
    \end{tabular}
    }
    \label{tab:mode}
    \vspace{-3pt}
\end{table*}

\subsection{Scalable Execution for Three Parallel Frameworks}
In this section, we discuss how our approach supports scalable execution via the three parallel frameworks in Section~\ref{sec:bigdata_background}, namely Spark-based, Dask-based, and Horovod-based analytics. The first two utilize virtual CPU clusters and the third utilizes virtual GPU clusters. By specifying the virtual machine type and number, cloud services can provision a cluster hardware environment. However, software dependencies, process coordination, and even access permission may differ for different big data analytics. Because of these differences, each framework requires its own \texttt{\small resources.json} and implementation of the first serverless function shown in~\figref{fig:timeline}. 
To reproduce big data analytics, one important part is to record and reuse original big data engine configurations. Paper~{\cite{perera2016reproducible}} uses separate files to record Spark memory configuration for reproducibility. Similar to this approach, we set these configurations by recording the information via command line arguments or original big data engine configuration files. Big data engine's configurations can be modified in application reproduction by users in \texttt{\small application.ini} file, such as changing \textit{--driver-memory 60g --executor-memory 60g} for Spark engine. Additional big data engine configurations are set up via separate files like spark-env.sh in the \textit{\$SPARK/conf} folder. Our toolkit supports storing such files in the cloud so they can be reused in reproduction. 

Beyond the listed frameworks, the additional parallel frameworks can also be deployed by updating the docker images' address in \texttt{\small application.ini}. RPAC will setup this parallel framework in the second serverless function of \figref{fig:timeline} and execute analytics within the new environment.

\neck{Spark-based big data analytics on virtual CPU nodes.}
We provide Spark-based parallel framework via the docker-based Spark engine virtual cluster provisioned by direct cloud services like AWS EMR with additional cloud resources like virtual network, container service and file system. 
By default setting, the resource manager like YARN NodeManager initiates the environment from a pulled docker image, and allocates one virtual instance as the master while others as workers. With serverless based pipeline execution, our toolkit enables automated execution management on master and execution computation on workers defined by serverless function handlers/implementations. 

Since big data analytics utilizes many compute nodes with complex computation proprieties, it is important to make sure availability and reliability during cloud execution. To achieve a secure and stable scalable execution, we control the access permission of master and workers by using the network security group. During big data analytics, our pipeline assigns one group for the master and another group for workers, and only enables TCP/UDP inbound and outbound rules within them. Also, for computation reliability, the big data analytics pipeline only allows client SSH permission for the master security group.

\neck{Dask-based big data analytics on virtual CPU nodes.}
Besides Spark, our RPAC toolkit also supports CPU-based parallel analytics by using Dask as the resource manager in the virtual cluster. Different from Spark which has dedicated cloud services (such as EMR in AWS), Dask environment can only be provisioned by regular virtual machine services (such as EC2 in AWS). 

Each virtual instance in the cluster initiates one docker container and our pipeline assigns one of the containers to be the Dask scheduler and others to be workers. Same with the security group setup with Spark-based analytics, we divide the client access between scheduler and workers for execution reliability. During execution, different from AWS EMR service which automatically initiates Spark processes after hardware provisioning, our RPAC toolkit needs to start Dask processes on both scheduler and worker containers during software provisioning before executing big data analytics on virtual CPU nodes. Besides, same as Spark-based cloud services, the client can also produce interactive visualizations based on Dask diagnostic dashboard in our framework, by using the public DNS name (public IP) of the scheduler instance with its dashboard port.

\neck{Horovod-based big data analytics on virtual GPU nodes.}
To provide a GPU-based parallel framework, we leverage Horovod and regular virtual machine services for analytics. The RPAC toolkit executes multi-instance GPU-based data analytics within our pre-built Docker containers, involving a shared file system and a customized port number for the SSH daemon. In order to categorize functionality between different instances, we set one of them as the primary worker and others as secondary workers. Within the container, the primary worker runs the MPI parallel command for data analytics execution while secondary workers listen to that specific port. 

\section{Reproduce Big Data Analytics in the Cloud}
\label{sec:reproduction}
In this section, we discuss how to achieve different levels of reproducibility within the same cloud and across different cloud providers. To achieve reproducibility, the user only needs to provide the URL of a historical execution stored in cloud storage (more in Section \ref{datamodeling}) and her own configurations. We will explain how our framework and RPAC toolkit support different ways of reproducibility summarized in Section~\ref{sec:reproduce_background}.

\subsection{Reproducibility in the Same Cloud} \label{sec:samecloud}

\neck{Reproduction with the same environment and configuration.}
This type of reproducibility is simplest because it is the same with the first execution as long as we can retrieve the information used from execution history. As illustrated by the first item in reproducibility phase of~\figref{fig:life-cycle}, by retrieving \texttt{\small resources.ini} and \texttt{\small application.ini} from execution history and providing proper \texttt{\small personal.ini}, our RPAC toolkit can rerun the experiment the same way it was executed for the first time. 

\neck{Reproduction in a different environment.}
Reproduction in a different environment means the virtual environment configuration needs to be changed from a historical execution, which is often useful for scale-up and scale-out experiments. As illustrated by the second reproducibility item in~\figref{fig:life-cycle}, a new \texttt{\small resources.ini} needs to be provided explaining the new environment setup (mostly virtual machine type and number). Then our RPAC toolkit can use it to generate a new executable \texttt{\small resources.json} and run the experiment in the same cloud. 

\neck{Reproduction with a different application configuration.} Reproduction with a different application configuration is often useful to run the application with a different dataset and/or application argument. As illustrated by the third reproducibility item in~\figref{fig:life-cycle}, a new \texttt{\small application.ini} needs to be provided explaining the new application setup. Then our toolkit can use it to generate a new executable \texttt{\small application.json} and run the experiment in the same cloud. 

We note that the last two reproductions can be easily combined for the requirements of running an application with different configurations and a different environment. To support it, a new \texttt{\small resources.ini} and a new \texttt{\small application.ini} should be provided.

\begin{table*}[ht]
\centering
\vspace{-3pt}
\caption{Cloud service used by reproducible and portable data analytics.} 
\vspace{-6pt}
\resizebox{.985\textwidth}{!}{%
\begin{tabular}{|c|c|c|c|c|}
\hline
\textbf{Service category} & \textbf{Service description} & \textbf{Amazon AWS} & \textbf{Microsoft Azure} & \textbf{Google Cloud}\\ \hline \hline
 \multirow{2}*{Virtual cluster} & Virtual machine cluster that enables to host distributed data & EC2 Auto Scaling & Virtual Machine Scale Set & Autoscaling Groups\\ 
 & analytics engines. & /EMR & /HDInsight & /Dataproc\\ \hline
 Virtual network & Manage and monitor networking functionality for cloud resources. & VPN & Virtual Network & Virtual Private Cloud\\ \hline
 Container service & Store, manage, and secure container images in private or public. & ECR & Azure Container Registry & Artifact Registry\\ \hline
 Object storage & Store, manage, and secure any amount of data in storage. & S3 & Blob storage & Firebase\\ \hline
 Database & Scalable and secure NoSQL cloud database. & DynamoDB & CosmosDB & Firebase Realtime Database\\ \hline
 \multirow{2}*{Serverless} & \multirow{2}*{Run and manage the application with zero server management.} & CloudFormation & Deployment Manager & Cloud Deployment Manager\\ 
 && \& Lambda Functions & \& Azure Functions & \& Cloud Functions\\\hline
 Cloud Python SDK & Easy-to-use interface to access cloud services. & Boto/Boto3 & .NET Core & Cloud SDK \\ \hline
 Authentication & Provide fine-grained access control for cloud resources. & AWS IAM & Azure IAM & Cloud IAM \\ \hline
\end{tabular}%
}
\vspace{-3pt}
\label{tab:extensibility}
\end{table*}

\begin{figure}[ht]
\removelatexerror
\vspace{-3pt}
\begin{algorithm}[H]
\small
\caption{\footnotesize Cloud~Agnostic~Application~Model~(CAAM)} \label{alg:caam}
\DontPrintSemicolon
\KwIn{resources.ini, application.ini, personal.ini}
\KwOut{pipeline.json}

  \SetKwFunction{FMain}{\textbf{function} CAAM}
  \SetKwFunction{FInit}{Init}
  \SetKwFunction{Faws}{\textbf{class} AWS}
  \SetKwFunction{Fazure}{\textbf{class} Azure}
  \SetKwFunction{FAdapter}{\textbf{class} CloudAdapter}
  \SetKwFunction{Mgen}{\textbf{method} generate}
  \SetKwFunction{MInit}{\textbf{method} $\_\_$Init$\_\_$}
  \SetKwFunction{Fraise}{\textbf{raise} NotImplementedError}
  \SetKwProg{Fn}{}{:}{}
  \;
  \Fn{\Faws}{
   \Fn{\MInit{$config$}}{
   $res \gets $\;
   $~~~~~~[config.res, config.para\_frame, ServMapping]$\\
   $resources.json \gets GetAwsRes(res)$\;
   $application.json \gets GetAwsApp(config.app)$\;
   $personal.json \gets $\;
   $~~~~GetAwsPersonal(config.personal)$\\
   }
   \Fn{\Mgen{}}{
   $pipeline \gets $\;
   $~~~~~~[resources.json, application.json, personal.json]$\\
   \KwRet $GetAwsPipeline(pipeline)$\;}
  }
  \;
  \Fn{\Fazure}{
   \Fn{\MInit{$config$}}{
   $res \gets $\;
   $~~~~~~[config.res, config.para\_frame, ServMapping]$\\
   $resources.json \gets GetAzureRes(res)$\;
   $application.json \gets GetAzureApp(config.app)$\;
   $personal.json \gets $\;
   $~~~~GetAzurePersonal(config.personal)$\\
   }
   \Fn{\Mgen{}}{
   $pipeline \gets $\;
   $~~~~~~[resources.json, application.json, personal.json]$\\
   \KwRet $GetAzurePipeline(pipeline)$\;}
  }
  \;
  \Fn{\FAdapter}{
    \Fn{\MInit{$AdaptedMethods$}}{
    $this.\_\_dict\_\_.update(AdaptedMethods)$
    }
  }
  \;
  \Fn{\FMain{}}{
    $AdapteeMapping \gets [aws:\texttt{AWS},azure:\texttt{Azure}]$\;
    $config \gets read($\textbf{Input}$)$\;
    \If{arg.reproduce}{
    $config.update(arg.reproduce)$\;
    }
    $cloud\_provider \gets config.cloud\_provider$\;
    \eIf{$cloud\_provider$ \textbf{in} $AdapteeMapping.keys$}{
    $Adaptee \gets AdapteeMapping[cloud\_provider]$\;
    $element \gets $\;
    $~~~~~(execute:Adaptee(config).generate)$\;
    $Pipeline \gets CloudAdapter(dict.add(element))$
    }{
    \textbf{raise} $NotImplementedError()$
    }
    \KwRet $Pipeline.execute()$ \KwTo \textbf{Output}\;
  }
\end{algorithm}
\vspace{-12pt}
\end{figure}

\subsection{Cross-cloud Reproducibility}
We discuss how the client achieves reproducibility with a different cloud provider. As illustrated in the fourth way of reproduction in \figref{fig:life-cycle}, by providing cloud service mapping and corresponding serverless function implementation, our toolkit can transform the general-purpose configurations in execution history into a new executable pipeline file for another cloud. 

To extend the reproducibility to another cloud, by leveraging the adapter pattern~\cite{w3s}, we propose a portable Cloud Agnostic Application Model (CAAM) in order to solve the vendor lock-in and interoperability problem for big data analytics, which is shown in Algorithm \ref{alg:caam}. When CAAM receives \texttt{\small resources.ini}, \texttt{\small application.ini} and proper \texttt{\small personal.ini}, \texttt{\small{CloudAdapter()}} invokes each vendor specific method of different cloud. It means as long as there is an adaptee class written for the cloud provider, by calling the \texttt{\small{CloudAdapter()}} with this cloud provider, the provided general-purpose configurations will be transformed to the executable request information of the target cloud based on its specification sets and execution requirements. By combining the compatible information of resources, application and personal, CAAM generates the overall executable \texttt{\small{pipeline.json}} and starts to execute the data analytics.

As shown in Algorithm \ref{alg:caam}, each cloud adaptee needs to implement how to get its~\texttt{\small{resources.json}}~based on \texttt{\small{resources.ini}} from execution history, parallel framework and service mapping shown in \tabref{tab:extensibility}. After all \texttt{\small{json}} files are ready, AWS uses \texttt{\small{GetAwsPipeline(AwsConfig)}} to generate pipeline file, while Azure uses \texttt{\small{GetAzurePipeline\\(AzureConfig)}} for generation. With CAAM, client directly calls \texttt{\small{CloudAdapter()}} with a specific adaptee method to execute data analytics with one general-purpose configuration. By calling \texttt{\small{Pipeline.execution()}}, the \texttt{\small{generate()}} method in corresponding cloud adaptee will generate the pipeline file and execute the big data analytics in cloud. Particularly, adaptee is in a modular design that can be injected into, removed from, or replaced within CAAM at any time.

\neck{Extensibility on cross-cloud reproduction.}
Our reproducible and portable big data analytics can be easily extended to additional clouds because most services from different cloud providers can be mapped to each other. \tabref{tab:extensibility} lists all cloud services provided by Amazon AWS, Microsoft Azure and Google Cloud for data analytics. Our toolkit currently only implements cross-cloud reproducibility between AWS and Azure. Extension to Google Cloud can be done by adding an additional cloud specific adaptee and providing corresponding service mapping with function implementation.

\section{Evaluation}
\label{sec:experiments}
We implement the reproducible and portable cloud computing and open-source it on GitHub at~\cite{Reproducible_and_Portable_Big_Data_Analytics_in_Cloud}. 
Two CPU-based analytics applications (cloud retrieval and causality discovery) and one GPU-based analytics application (domain adaptation) are tested in our experiments. All benchmark evaluations are developed on two cloud providers, Amazon AWS and Microsoft Azure. Seven metrics are used to evaluate our work which include data analytics metrics like execution time, budgetary cost, cost-performance ratio, and cloud reproducibility metrics like overhead. 

\begin{table}[h]
    \centering
    \small
    \vspace{-3pt}
    \caption{Comparison of cloud resources.}
    \vspace{-6pt}
    \resizebox{.49\textwidth}{!}{
    \begin{tabular}{|c|c|c|c|c|c|}
    \hline
        \textbf{Framework}&\textbf{Metrics}&\textbf{Cloud}&\textbf{Type}&\textbf{vCPU}&\textbf{Memory (GiB)}\\
        \hline
        \hline
        \multirow{4}*{CPU-based} & \multirow{2}*{scale-up} & AWS & c5d.4xlarge & 16 & 32 \\
        \cline{3-6} 
        && Azure & F16s$\_$v2 & 16 & 32\\
        \cline{2-6}
        & \multirow{2}*{scale-out} & AWS & c5d.(x)large & 2 (4) & 4 (8) \\
        \cline{3-6} 
        &&Azure& Fs$\_$v2 & 2 & 4 \\
        \hline
        \multirow{4}*{GPU-based} & \multirow{2}*{scale-up} & AWS &  p3.8xlarge & 4 & 16 \\
        \cline{3-6} 
        && Azure & NC24s$\_$v3 & 4 & 16 \\
        \cline{2-6} 
        & \multirow{2}*{scale-out} & AWS & p3.2xlarge & 1 & 16 \\
        \cline{3-6} 
        &&Azure& NC6s$\_$v3 & 1 & 16 \\
        \hline
    \end{tabular}
    }
    \label{tab:resources}
    \vspace{-6pt}
\end{table}

\tabref{tab:resources} lists the exact cloud resources we use for each data analytics. For executing the application with a larger dataset, additional storage like AWS Elastic Block Store (EBS) is also been attached during the resource initialization.
One variation is in scale-out of AWS CPU-based evaluation. We use c5d.large cluster for Dask-based analytics, but c5d.xlarge cluster for Spark-based analytics because AWS EMR requires more computational capability.

\subsection{Benchmark Analytics and Datasets}
To benchmark our toolkit's functionality comprehensively, we employ four applications and each uses a separate big data framework in Section~\ref{sec:bigdata_background}.

\neck{Cloud retrieval.} Cloud property retrieval is an important task in remote sensing and Atmospheric science. We used the implementation of paper~\cite{wang2020machine} for our first application. It trains a Random Forest machine learning model for cloud mask and cloud thermodynamic-phase retrieval from satellite observations. Dask framework is used for execution parallelization. The Docker image we built is hosted on DockerHub public repository, with Python 3.6 and sklearn 0.24.2. Total datasets are around 0.5 GB. 

\neck{Causality discovery.} In order to discover the cause-effect relationships in a system with the increasing volume and dimensionality of available data, the two-phase scalable and hybrid causality discovery is proposed by Guo et al.~\cite{9288491}. As a big data analytics, we use the Spark application with Hadoop in the cloud virtual cluster. The Docker image we built is hosted on DockerHub public repository, with Python 3.7 and R 3.4. The data in our execution is 200,000 rows of simulated five variable time-series records, which is around 10 MB.

\neck{Domain adaptation.} Unsupervised Domain Adaptation (UDA) aims to transfer the knowledge learned from a labeled source domain to an unlabeled target domain. We use the UDA implementation designed by Sun et al.~\cite{sun2016deep} that solves the problem of the unlabeled target domain. To move this data analytics to the cloud, we use the virtual cluster with Pytorch GPU acceleration and Horovod with MPI. The Docker image we built is hosted on DockerHub public repository, with Python 3.6, CUDA 10.1 and cuDNN 7. The data we used is the public Office dataset containing 31 object categories in two domains: Amazon and Webcam, which is around 50 MB in total.

{\neck{Satellite collocation.}} Because there are many satellites orbiting the Earth, it is valuable to integrate and/or compare their measurements. Satellite collocation provides a way to pair measurements from two satellite sensors that observe the same location quasi-simultaneously. We implemented and parallelized the method in~{\cite{Holz2008}} to generate collocated data from two satellites. Like the cloud retrieval application, we use Dask framework for execution parallelization. The Docker image we built is hosted on DockerHub public repository, with Python 3.8, Pandas 1.5.0 and H5py 3.7.0. The two satellites we used in the experiment include the ABI passive sensing data product from NOAA Geostationary Operational Environmental Satellites (GOES-16+)~{\cite{noaa-goes}} and the CALIOP active sensing data product from NASA Cloud-Aerosol Lidar and Infrared Pathfinder Satellite Observations (CALIPSO) satellite~{\cite{winker2009overview}}. The total input data volume is 1.1 TB.

\subsection{Evaluation Metrics}
Even though there have been many studies on reproducibility, as stated in this recent survey paper~\cite{ivie2018reproducibility}, there are still no agreed metrics that can quantitatively measure reproducibility and compare different reproducible toolkits. The survey paper thinks performance, scalability and efficiency are possible metrics, but no concrete metric definition was provided. In this work, to promote fair comparison, we provide our own definition of performance, scalability and efficiency for cloud based reproducibility, which results in seven metrics (namely \textit{m1} to \textit{m7} listed below). 

\subsubsection{Execution Performance Metrics} 
Following paper~\cite{li2017quantitative}, we measure execution performance of our cloud based application via \textit{m1: execution time}, \textit{m2: budgetary cost} and \textit{m3: performance-price ratio (PPR)}.
We first record execution time for each data analytics benchmark. The execution time is the wall-clock time of analytics pipeline (as shown in \figref{fig:timeline}), which includes pipeline file preparation, cloud resources deployment and initialization, data analytics execution, execution history upload, and termination.

Budgetary cost contains bill usages for all resources used in each data analytics benchmark, which mainly includes the virtual cluster, container, network, database, and object storage with read and write request usage.

Regarding the performance-price ratio (PPR), it evaluates the performance of each analytics considering the execution time with cost. We use the same formula used in~\cite{wang2014workflow} for PPR by calculating the product of execution time and budgetary cost. Lower PPR is more desirable excluding other factors. 

\begin{figure*}[b]
	\centering
	\vspace{-6pt}
	\subfloat[Execution time.]{
		\input{figure/CR_time}
		\label{fig:CR_time}
	}
	\subfloat[Budgetary cost.]{
		\input{figure/CR_cost}
		\label{fig:CR_cost}
	}
	\subfloat[Performance-price ratio.]{
		\input{figure/CR_ratio}
		\label{fig:CR_ratio}
	}
	\vspace{-3pt}
	\caption{Scalability evaluation of RPAC toolkit for the \textbf{cloud retrieval application}: \textbf{scale-up} (circle and square mark) and \textbf{scale-out} (triangle and diamond mark) for AWS and Azure. Dashed line: cost value calculated by its usage.}
	\label{fig:CR_scala_up_out}
	\subfloat[Execution time.]{
		\input{figure/CD_time}
		\label{fig:CD_time}
	}
 	\subfloat[Budgetary cost.]{
 		\input{figure/CD_cost}
		\label{fig:CD_cost}
	}
 	\subfloat[Performance-price ratio.]{
		\input{figure/CD_ratio}
 		\label{fig:CD_ratio}
	}
	\vspace{-3pt}
	\caption{Scalability evaluation of RPAC toolkit for the \textbf{causality discovery application}: \textbf{scale-up} (circle mark) and \textbf{scale-out} (triangle mark) for AWS. Dashed line: cost value calculated by its usage.}
 	\label{fig:CD_aws}
 	\vspace{-3pt}
 \end{figure*}

\subsubsection{Cloud Scalability Metrics} We evaluate the scalability of our work for both \textit{m4: scale-up (vertical scaling)} and \textit{m5: scale-out (horizontal scaling)}.

Cloud scale-up is achieved by utilizing more resources within an existing computation system to reach a desired state of performance. In our evaluation, scale-up is set in a single virtual machine by having more threads in Dask-based analytics, more executor cores in Spark-based analytics, or more GPUs in Horovod-based analytics. For cloud retrieval, we fix the number of threads for each Dask worker, and utilize the number of workers from 1 to 8 during evaluation. It is the same in domain adaptation, except by increasing more threads for GPUs rather than CPUs. For causality discovery, because of the EMR setup, we launch only one worker in each virtual instance and allocate only one executor in this worker. To scale up, we use one executor with increasing the numbers of vCPUs of this executor for parallel execution.

In real world scale-up, it is undesired to launch a powerful instance but only use its partial computational capability. In order to have a fair comparison, we additionally measure scale-up cost by usage, which times the budgetary cost of one instance by the percentage of CPU that is actually used. It simulates scale-up scenarios that use more and more powerful machines.

Cloud scale-out is usually associated with a distributed architecture, which is achieved by adding additional computational capacity to a cluster. In our evaluation, scale-out is set by increasing more virtual machines in an existing cluster. For cloud retrieval and domain adaptation, we deploy only one worker process per instance, and increase the number of instances from 1 to 8 during evaluation. For causality discovery, we instead use one CPU core in each executor, and increase the number of workers by adding virtual instances. 

\subsubsection{Reproducibility Efficiency Metrics}
For reproducibility, a metric \textit{m6: reproducibility\_overhead} is used to understand how much overhead it brings by supporting reproducibility during execution. Since reproducibility support is achieved by storing application configuration and execution history, we calculate the ratio between additional execution time caused by reproducibility data storage and the execution time of execution without reproducibility support. The lower the overhead ratio is, the better. 

As we mentioned in Section \ref{sec:serverless_automation}, an SDK-based pipeline execution mode has also been proposed for big data analytics. Since both SDK-based and serverless-based approaches can be achieved automatically, we also measure \textit{m7: reproducibility\_efficiency} to compare their execution time with reproducibility.

\begin{figure*}[t]
	\centering
	\vspace{-6pt}
	\subfloat[Execution time.]{
		\input{figure/DA_time}
		\label{fig:DA_time}
	}
	\subfloat[Budgetary cost.]{
		\input{figure/DA_cost}
		\label{fig:DA_cost}
	}
	\subfloat[Performance-price ratio.]{
		\input{figure/DA_ratio}
		\label{fig:DA_ratio}
	}
	\vspace{-3pt}
	\caption{Scalability evaluation of RPAC toolkit for the \textbf{domain adaptation application}: \textbf{scale-up} (circle and square mark) and \textbf{scale-out} (triangle and diamond mark) for AWS and Azure. Dashed line: cost value calculated by its usage.}
	\label{fig:DA_scala_out}

	\subfloat[Execution time.]{
		\input{figure/SC_time}
		\label{fig:SC_time}
	}
	\subfloat[Budgetary cost.]{
		\input{figure/SC_cost}
		\label{fig:SC_cost}
	}
	\subfloat[Performance-price ratio.]{
		\input{figure/SC_ratio}
		\label{fig:SC_ratio}
	}
	\vspace{-3pt}
	\caption{Scalability evaluation of RPAC toolkit for the \textbf{satellite collocation application}: \textbf{scale-up} (circle mark) and \textbf{scale-out} (triangle mark) for AWS. Dashed line: cost value calculated by its usage.}
	\label{fig:SC_scala_up_out}	
	\vspace{-6pt}
\end{figure*}

\subsection{Benchmarking for Execution Performance and Scalability}
In this section, we first assess the cloud scalability of our RPAC toolkit based on three metrics: \textit{m1: execution time}, \textit{m2: budgetary cost}, and \textit{m3: performance-price ratio}. In \textit{m4: scale-up} evaluation, the execution is analyzed by gradually utilizing more resources in one instance. In \textit{m5: scale-out}, the evaluation is achieved by gradually adding additional instances of the same type in one cluster. 
Next, we will explain our benchmarking results of the four applications. 

\subsubsection{Scalability Evaluation for the Cloud Retrieval Application}
The cloud retrieval \textit{m4: scale-up} and \textit{m5: scale-out} evaluations are shown in \figref{fig:CR_scala_up_out}. 
As illustrated in \figref{fig:CR_time}, the \textit{m1: execution time} decreases when the number of executors increases in both AWS and Azure with similar trends. The \textit{m2: budgetary cost} as shown in \figref{fig:CR_cost}, however, decreases in \textit{m4: scale-up} and increases in \textit{m5: scale-out} when the number of executors increase. The reason is that in cluster scale-up, the same resources were used while their execution time was decreasing; and in cluster scale-out, the costs saved by less execution time costs were less than the costs increased with additional resources. If only calculating the cost by usage for \textit{m4: scale-up} case, its trends become similar to those of \textit{m5: scale-out}. Combining cost and time, as illustrated in \figref{fig:CR_ratio}, the \textit{m3: PPR} in \textit{m4: scale-up} and scale-up by usage decrease when the numbers of executors increase. However, the \textit{m3: PPR} first decreases but later increases a little bit in \textit{m5: scale-out} cases. 
The figure also shows AWS achieves better \textit{m3: PPR} than Azure, and \textit{m4: scale-up} achieves better \textit{m3: PPR} than \textit{m5: scale-out}. So the best \textit{m3: PPR} for the Dask-based big data application with virtual CPU nodes is achieved by \textit{m4: scale-up} of application with more executors in AWS.

\subsubsection{Scalability Evaluation for the Causality Discovery Application}
Because Azure HD-Insight cluster does not support Docker-based Spark computation, we only focus on the evaluation of causality discovery for AWS, which is shown in \figref{fig:CD_aws}. The trends for this application are very similar to those for the previous application since they both are CPU-based. As illustrated in {\figref{fig:CD_time}}, the \textit{m1: execution time} for both \textit{m4: scale-up} and \textit{m5: scale-out} decreases dramatically by at most 80\% when the parallelism increases. This change of time appears more significant in causality discovery compared with what is in cloud retrieval. 
For the \textit{m2: budgetary cost} in {\figref{fig:CD_cost}}, when the parallelism increases, the \textit{m4: scale-up} decreases, while both \textit{m5: scale-out} and scale-up by usage increase with similar trends. 
For all three metrics in {\figref{fig:CD_ratio}}, The \textit{m3: PPR} decreases when the numbers of executors increase. As a result, it is better to use a larger number of executors in the Spark-based big data analytics with virtual CPU nodes.

\subsubsection{Scalability Evaluation for the Domain Adaptation Application}
For domain adaptation, the evaluations are shown in \figref{fig:DA_scala_out}. Because the maximal number of GPUs in one instance is 4 for Azure, we compare \textit{m4: scale-up} only from 1 GPU to 4 GPUs. As illustrated in \figref{fig:DA_time}, same with the findings from other data analytics, the \textit{m1: execution time} decreases when the numbers of GPUs increase in both AWS and Azure. The \textit{m2: budgetary cost} in \figref{fig:DA_cost}, also have the same regularity compared with CPU-based analytics. For \textit{m3: PPR},  as illustrated in \figref{fig:DA_ratio}, more GPUs lead to better ratios for \textit{m4: scale-up} and worse ratios for scale-up by usage. For \textit{m5: scale-out}, \textit{m3: PPR} first gets worse and then improves a little bit. But still launching with only 1 instance can have the best \textit{m3: PPR} for both AWS and Azure execution. 

\subsubsection{Scalability Evaluation for the Satellite Collocation Application}
The above three applications already show the effectiveness of RPAC for parallel frameworks in different clouds. 
We further evaluate the satellite collection application with over 1 TB input data on AWS, and its longest total execution time is over 25 hours. As shown in {\figref{fig:SC_scala_up_out}}, the \textit{m1: execution time} in {\figref{fig:SC_time}} of all \textit{m4: scale-up} experiments decrease around 1 to 2 hours compared with all \textit{m5: scale-out} experiments in the same parallelism setting. Thus, parallel execution in one VM with scale-up deployment is preferred, since \textit{m5: scale-out} generates more communication overheads between different nodes. For the \textit{m2: budgetary cost} as illustrated in {\figref{fig:SC_cost}}, when the number of executors increases, the \textit{m4: scale-up} gets a more reasonable price while the \textit{m5: scale-out} becomes more expensive. Different with previous findings, \textit{m2: budgetary cost} of \textit{m5: scale-out}, \textit{m4: scale-up} and scale-up by usage change very dramatically by at most 75\% when the parallelism changes. The reason is that the execution time of the big data application is much longer than others. Combining cost and time, as illustrated in {\figref{fig:SC_ratio}}, the \textit{m3: PPR} of \textit{m4: scale-up} is decrease when the numbers of executors increase. The \textit{m3: PPR} of scale-up by usage and \textit{m5: scale-out} are first decrease but later increase a little bit. As a result, the better parallelism strategy for the big data application is using more executors in \textit{m4: scale-up} deployment.

\subsection{Benchmarking for Reproducibility Efficiency}
In this section, we assess the efficiency of reproducibility for RPAC toolkit in the first three applications. We first evaluate the overhead caused by serverless-based reproducibility, then we compared the efficiency between serverless-based and SDK-based approaches.

\begin{figure}[b]
	\centering
	\vspace{-3pt}
	\subfloat[The box-plots.]{
		\input{figure/CR_boxplot}
	}
	\hspace{1pt}
	\subfloat[The relative difference.]{
		\input{figure/CR_boxplot_relative}
		\label{fig:relative1}
	}
	\vspace{-3pt}
	\caption{The box-plots and its relative difference for application execution time with and without reproducibility support.}
	\label{fig:box-plot}
	\vspace{-3pt}
\end{figure}
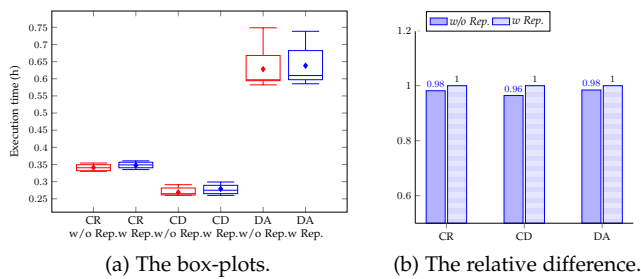

\subsubsection{Efficiency Comparison for Reproducibility Support} We first measure the \textit{m6: reproducibility\_overhead} of our applications with and without reproducibility support. For each application, we measure the AWS scale-up with 4 parallelisms, run each experiment 10 times and collect all results in a box-plot shown in \figref{fig:box-plot}. From the figure, we can see having reproducibility support did not cause much overhead, which is less than $0.01$ hours, for all applications. The overhead percentage caused by reproducibility for cloud retrieval (CR), causality discovery (CD), and domain adaptation (DA), are 1.28\%, 3.58\%, and 2.17\%, respectively. Besides, the time range of GPU-based analytics is larger than both CPU-based analytics, which means the execution time of GPU-based computation is more unstable than CPU-based one.

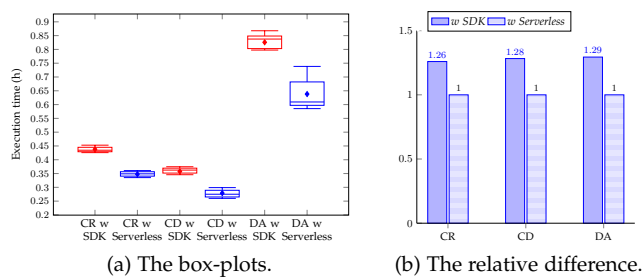
\begin{figure}[b]
	\centering
	\vspace{-3pt}
	\subfloat[The box-plots.]{
		\input{figure/SDK_boxplot}
	}
	\hspace{1pt}
	\subfloat[The relative difference.]{
		\input{figure/SDK_boxplot_relative}
		\label{fig:relative2}
	}
	\vspace{-3pt}
	\caption{The box-plots and its relative difference for application execution time with serverless-based and SDK-based approach}
	\vspace{-3pt}
	\label{fig:box-plot2}
\end{figure}

We utilize a statistical hypothesis test approach, called T-test~\cite{kim2015t}, to determine whether the execution time with and without reproducibility support differ statistically. T-test determines a possible conclusion from two different hypotheses. By calculating the corresponding p-value~\cite{dahiru2008p}, we can measure the probability that an observed difference has occurred just by random chance. Hypothesizing that the reproduce execution provides some overhead over the execution without reproducibility, we calculate the p-value for the two sample t-test with equal variance. The p-values of CR, CD and DA, turn out to be 0.4968, 0.3193 and 0.3634. Since these are not less than $p = 0.05$, we fail to reject the null hypothesis of the tests. As the result, we do not have sufficient evidence to say that the average execution time between the two species (with and without reproducibility support) is different for all three applications.

\subsubsection{Comparison with SDK-based Reproduction} Besides the serverless-based approach, as shown in \tabref{tab:mode}, we also implemented an SDK-based automatic execution mode which is achieved by periodical status pulling. In order to explore their difference, we evaluate the \textit{m7: reproducibility\_efficiency} of these two approaches with the same applications. Same with the previous measurement setting, we run each experiment 10 times and collect all results in a box-plot as illustrated in \figref{fig:box-plot2}. For SDK-based approach, the time window for each status pulling is set to 10 second. The figure shows that serverless-based approach is more efficient than SDK-based approach, and the percentage of overhead reduction for CR, CD and DA, are 25.92\%, 28.24\% and 29.41\%, respectively. The time range of serverless-based approach is larger than SDK-based one especially in GPU-based analytics. The reason is that, in SDK-based approach, the execution status monitoring could be delayed with periodical pulling. With the serverless function and event trigger, serverless-based approach enables big data analytics to be measured more efficiently and with less noise.

We also use T-test to determine whether the execution time using serverless approach and SDK-based approach differs  statistically. Hypothesize that the serverless-based approach provides some efficient benefit over SDK-based approach. The p-values of CR, CD and DA, turn out to be $9.31\mathrm{e}{-15}$, $6.00\mathrm{e}{-14}$ and $2.19\mathrm{e}{-06}$. Since these p-values are less than $0.05$, we can reject the null hypothesis of the tests. The serverless-based approach is indeed providing statistically significant efficient benefit compared with SDK-based approach.

\begin{table*}[t]
\centering
\vspace{-1pt}
\footnotesize
\caption{Comparison of related work for cloud-based reproducibility.}
\vspace{-3pt}
\resizebox{.98\textwidth}{!}{%
\begin{tabular}{|c|c|c|c|c|}
\hline
\multirow{2}*{\textbf{Approach}}  &  \textbf{Scalable}  & \textbf{Automated execution} & \textbf{History retrieval} & \textbf{Cross-cloud}\\
& \textbf{environment provision} & \textbf{and reproducibility} & \textbf{based reproducibility} & \textbf{reproducibility}\\
\hline
\hline
ReproZip~\cite{chirigati2016reprozip}, CARE~\cite{janin2014care}, DevOps-based~\cite{boettiger2015introduction}, & \multirow{2}*{\LEFTcircle} & \multirow{2}*{\Circle} & \multirow{2}*{\Circle} & \multirow{2}*{\Circle} \\
Skyport~\cite{gerlach2014skyport}, Hyperflow~\cite{orzechowski2020reproducibility}, TOSCA~\cite{binz2014tosca,qasha2017automatic} &&&& \\
\hdashline[1pt/1pt]
AMOS~\cite{strijkers2011toward}, WSSE~\cite{dudley2010silico}, PDIFF~\cite{pdiff}, ReCAP~\cite{ahmad2016scientific}, Tapis~\cite{stubbs2021tapis} & \multirow{2}*{\CIRCLE} & \multirow{2}*{\Circle}  & \multirow{2}*{\CIRCLE} & \multirow{2}*{\Circle}  \\
OpenWhisk~\cite{openwhisk,baldini2016cloud}, NeuroCAAS~\cite{abe2022neuroscience} & & & & \\
\hdashline[1pt/1pt]
PRECIPE~\cite{azarnoosh2013introducing}, Chef~\cite{klinginsmith2011towards}, Apt~\cite{ricci2015apt}, Semantic Driven~\cite{santana2014semantic, santana2014leveraging} & \CIRCLE & \Circle & \Circle & \CIRCLE \\
\hdashline[1pt/1pt]
\textbf{RPAC} (this work) & \CIRCLE & \CIRCLE & \CIRCLE & \CIRCLE \\
\hline
\end{tabular}
}
\label{tab:related}
\vspace{-3pt}
\end{table*}

\section{Related Work} 
\label{sec:relatedworks}

There have been many studies on cloud-based reproducibility. Some of them~\cite{boettiger2015introduction,strijkers2011toward, dudley2010silico, pdiff, ahmad2016scientific} only study its conceptual frameworks. In this section, we only discuss those having actual systems/toolkits. As shown in \tabref{tab:related}, we categorize related work into four groups based on their systems' capabilities. 
Besides, we also selected two most related works to compare in detail. The comparison is shown in {\tabref{tab:related_most}} where the first one also leverages serverless computing and the second is one of the most recent work on cloud based reproducibility.

\begin{table*}[b]
\centering
\vspace{-1pt}
\caption{Detailed comparison among serverless computing related work.}
\footnotesize
\vspace{-3pt}
\resizebox{.995\textwidth}{!}{%
\begin{tabular}{|c|c|c|c|c|}
\hline
\multirow{2}*{} & \multirow{2}*{\textbf{Pipeline description}} & \textbf{Automated deployment,} & \textbf{Parallel execution} & \textbf{Capability of cross-cloud}\\
& & \textbf{execution and reproduction} & \textbf{on cloud} & \textbf{deployment/reproducibility}\\
\hline
\hline
\multirow{3}*{OpenWhisk~\cite{openwhisk,baldini2016cloud}} & Consists of events, triggers and & User needs to initiate the cloud cluster, & No explicit parallel framework support. & Not supported directly. \\
& action functions, which are & then provides the address of Kubernetes & Users have to provide environment and & Users have to rewrite the pipeline \\
& implemented by themselves. & to Openwhisk. No resource termination. & implement the pipeline by themselves. & by themselves. \\
\hdashline[1pt/1pt]
\multirow{3}*{NeuroCAAS~\cite{abe2022neuroscience}} & Consists of the specification of analysis & Pull pipeline from public repository and & No explicit parallel framework support. & Not supported directly. \\
& and infrastructure stack. Toolkit provides & select listed configurations. The options & Could be scripted and implemented & Users have to rewrite the pipeline \\
& formatted pipelines in a public repository. & are limited. No resource termination. & by users in the pipeline. & by themselves.  \\
\hdashline[1pt/1pt]
\multirow{3}*{\textbf{RPAC}} & Consists three aspects of abstraction & Full automated Cloud SDK mode and & Provide three parallel frameworks for Spark & By modifying three aspects \\
& and the serverless functions. Users can & serverless mode. Include hardware & -based and Dask-based analytics on CPUs, & of abstraction, RPAC enables \\
& update configurations from templates. & provisioning and resource termination. & and Horovod-based analytics on GPUs. & different levels of reproduction. \\
\hline
\end{tabular}
}
\vspace{-1pt}
\label{tab:related_most}
\end{table*}

\subsection{General Comparison with Related Work}

Among the related studies in \tabref{tab:related}, nearly all related approaches achieve the software environment provision for reproducibility. However, approaches in group 1 mainly use the archived or containerized software environment, which limits the scope of applicability and lacks support for maintaining hardware configurations within cloud. Additionally, they monitor execution status by system commands or cloud APIs based periodical pulling which is less efficient than event-based execution in our work. For example, ReproZip~\cite{chirigati2016reprozip} tracks system commands and zips collected information along with all the used system files together for reproducibility. CARE~\cite{janin2014care} reproduces a job execution by monitoring and archiving all the material required to re-execute operations. For related work in group 2, their proposed approaches encapsulate the code dependencies and software in virtual machine images or graphs, and enable history retrieval for reproduction. For instance, WSSE~\cite{dudley2010silico} proposes to generate digital data and source code snapshots to be reproduced and distributed within a cloud-computing provider. The Tapis~{\cite{stubbs2021tapis}} open-source API platform was proposed for accomplishing distributed computational experiments in a secure, scalable, and reproducible way. With the implemented pipeline with the Python API, the containerized applications can be submitted, scheduled and executed as tasks using a traditional HPC batch scheduler such as SLURM. AMOS~\cite{strijkers2011toward} uses a VM containing a set of tools previously installed to implement a mechanism that initializes and configures VMs on demand. However, this reproduction is more like a history repetition, which is designed for verification and validation of history execution. They also provide configurable environment variables for automatic resource deployment in a single cloud, but do not support cross-cloud reproducibility. Instead, our proposed RPAC uses a data abstraction for information needed for reproducibility and transforms resource configurations used in one cloud into those in another cloud. 

For related studies in \tabref{tab:related}, group 3's capabilities are closest to ours. These approaches rely on annotated information provided by a user to assign workflow, and software/hardware environment. For example, PRECIPE~\cite{azarnoosh2013introducing} provides APIs to access both AWS and private cloud. However, users need to call the functions in order and have to manually terminate resources after the experiment is done, so it does not support automated end-to-end execution and reproducibility. The whole execution has to wait at the client side before the next function can be called. On the contrary, RPAC serverless event triggering enables fewer communications between client and cloud, which improves the efficiency for cloud analytics. Chef~\cite{klinginsmith2011towards} achieves virtual execution environment launching and termination via designed knife commands. Chef client is installed in virtual machines to run the pipeline within the virtual machines. So some internal steps of the application can be executed within the virtual machines via its pipeline. However, Chef does not support full automation since its user has to wait at client side to manually terminate resources after the experiment is done. Apt~\cite{ricci2015apt} uses user-provided profiles, which consists of a cluster with a control system to instantiate encapsulated experiment environments, for repeating historical research. From this information, they deduce the required execution resources in cloud and then re-provision or configure them through their own APIs. In comparison, we use a serverless-based pipeline and follow cloud function APIs provided by cloud providers so the execution/reproduction process can be managed by the cloud without communications with toolkit. Also, their fully created annotations, even in cross-cloud reproduction, rely heavily on the users instead of execution history. Our work abstracts information required by users from information in execution history, users only need to provide minimal information to reproduce. Our toolkit will transform user-provided information into executable pipeline. For automated execution and reproducibility, none of these approaches can achieve full automation including resources and software provisioning, analytics execution and termination. RPAC's event-based automation is the one-command execution that achieves a more efficient cloud computation and reproduction. For cross-cloud reproducibility, we further use the adapter pattern model to achieve the configuration mapping without taking all inputs from the user.

\subsection{Detailed Comparison with Most Related Work}
As shown in {\tabref{tab:related_most}}, Apache OpenWhisk~{\cite{openwhisk}\cite{baldini2016cloud}} is an open-source, distributed Serverless cloud platform. In their serverless design, functions are explicitly defined in terms of the event, trigger, and action, which are implemented by users. Events are generated from event sources, which often indicate changes in data or carry data themselves. The trigger is defined by specifying its name and parameters (key-value pairs). It is associated with an action. The action is defined as functions (code snippets), which encapsulate application logic to be executed in response to events. Before deployment, users need to initiate the cluster in the cloud and provide the hostname and port of the Kubernetes cluster to the toolkit. No resource termination option when deployment finishes. Openwhisk has three deployment options. 1) OpenWhisk can be deployed using Helm charts on any Kubernetes provisioned from a public cloud provider. 2) The deployment can be achieved by OpenWhisk REST API or OpenWhisk CLI. 3) Use the cloud-defined CLI on a cloud provider that already provisions Apache OpenWhisk as a service, which is only supported by IBM cloud as of now. For parallel execution, Openwhisk does not provide direct parallel framework support. To enable scalable execution, users need to initiate a cloud cluster with the parallel software environment and prepare docker images with a parallel framework. Then users need to implement the parallel logic in the pipeline's action. OpenWhisk enables the deployment on different clouds with Helm charts on any Kubernetes. However, an application for one cloud cannot be redeployed in another cloud, unless users 1) initiate instances on another cloud with the Kubernetes cluster, 2) rewrite the events and triggers in the pipeline, and 3) provide the new hostname and port of Kubernetes cluster to the toolkit and redeploy the pipeline.
Another example is NeuroCAAS~{\cite{abe2022neuroscience}}. NeuroCAAS provides formatted pipelines, called blueprints, in a public code repository and defines a resource bank that can make hardware available through pre-specified instances in one specific cloud. The users are able to update the blueprint with new configurations and upload its new version to the public repository for deployment and reproduction. The users need to provide the blueprint's repository address for automated deployment, execution and reproduction. By default, NeuroCAAS fixes a single instance type per analysis in order to facilitate reproducibility. With the blueprint, datasets, and configuration files for one analysis, NeuroCAAS achieves reproducibility for corresponding analyses with the same environment and configuration. For parallel execution, NeuroCAAS does not provide direct parallel framework support. The logic of parallel processing must be explicitly scripted and implemented in the blueprint by users.

\section{Conclusions} \label{sec:conclusions}
Reproducibility is an important way to gain the confidence of new research contributions. In this paper, we study how to achieve cloud-based reproducibility for big data analytics. By leveraging serverless, containerization and adapter design pattern techniques, our proposed approach and RPAC toolkit can achieve reproducibility, portability and scalability for big data analytics. Our experiments show our toolkit can achieve good scalability and low overhead for reproducibility support for both AWS and Azure.

For future work, we will mainly focus on the following three aspects. First, we will optimize the executions in terms of time, cost or ratio by mining execution history, and further optimize the overhead of reproducibility via better data abstraction, modeling and storage.
Second, we will extend our work to easily publish data analytics as public records following the Research Object framework~\cite{bechhofer2013linked} so they can be referred via DOI identifiers later. Third, we will study how to utilize execution history data to achieve automated execution optimization based on users' objectives (time, cost or ratio) and datasets.

\section*{Acknowledgment}
This work is supported by the National Science Foundation (NSF) Grant No. OAC--1942714, National Aeronautics and Space Administration (NASA) grant No. 80NSSC21M0027 and U.S. Army Grant No. W911NF2120076.

\ifCLASSOPTIONcaptionsoff
  \newpage
\fi

\bibliographystyle{IEEEtran}
\bibliography{IEEEabrv,ref_nomark}

 \vskip -2\baselineskip plus -1fil 
\begin{IEEEbiography}[{\includegraphics[width=1in,height=1.25in,clip,keepaspectratio]{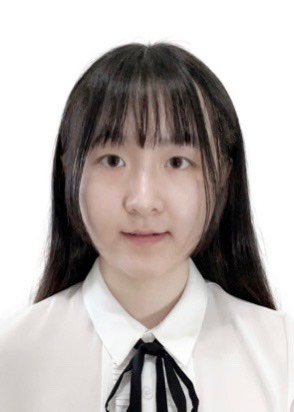}}]{Xin Wang}
	 received her PhD in Information Systems from the University of Maryland, Baltimore County in 2022. Her PhD research interests include distributed computing (systems), blockchains, big data analytics, federated learning, cloud computing and reproducibility.
\end{IEEEbiography}

 \vskip -2\baselineskip plus -1fil 
\begin{IEEEbiography}[{\includegraphics[width=1in,height=1.25in,clip,keepaspectratio]{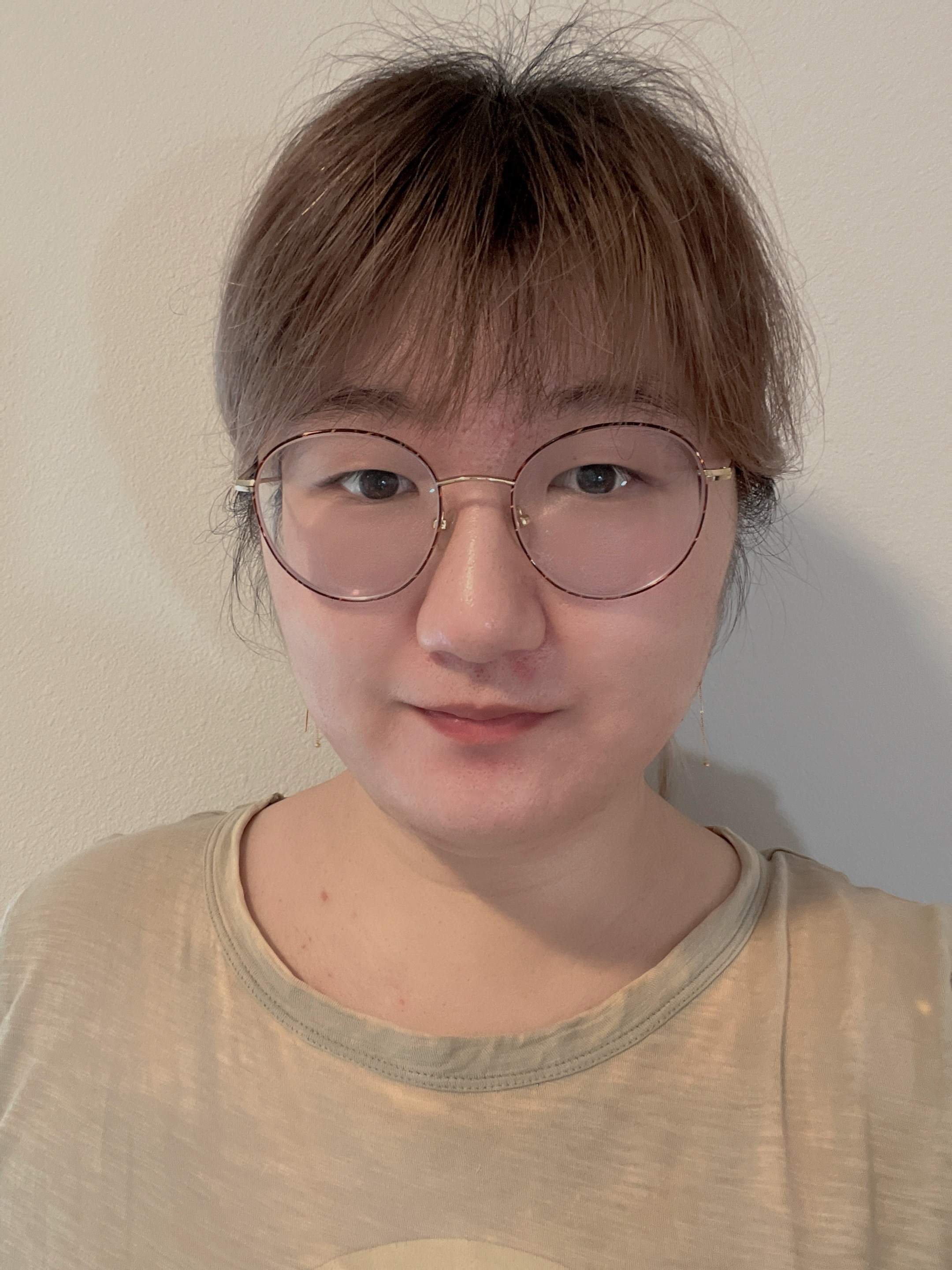}}]{Pei Guo}
	 is working as a Data Scientist at Wyze Labs. She received her PhD in Information Systems from the University of Maryland, Baltimore County in 2021. Her PhD researches focused on spatiotemporal causal modeling on large-scale data, big data application parallelizing and cloud computing. Her contribution to the paper was done during her PhD study at UMBC.
\end{IEEEbiography}

 \vskip -2\baselineskip plus -1fil 
\begin{IEEEbiography}[{\includegraphics[width=1in,height=1.25in,clip,keepaspectratio]{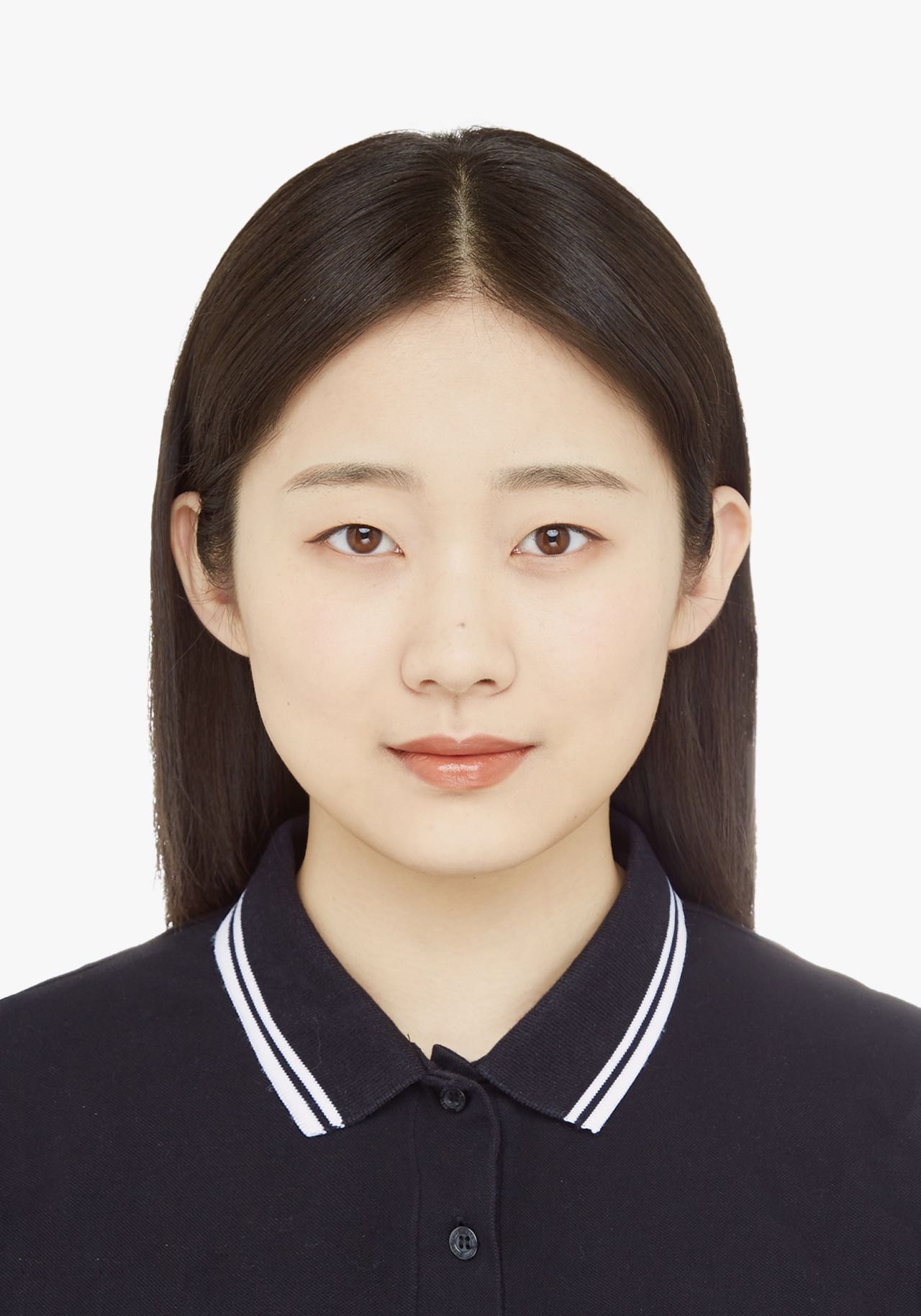}}]{Xingyan Li}
	 is a PhD student working in the Big Data Analytics Lab at the Department of Information Systems, University of Maryland, Baltimore County. Her research interests include data science, machine learning and deep learning.
\end{IEEEbiography}

 \vskip -2\baselineskip plus -1fil 
\begin{IEEEbiography}[{\includegraphics[width=1in,height=1.25in,clip,keepaspectratio]{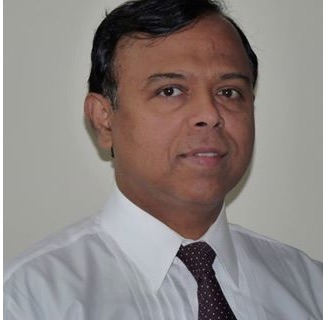}}]{Aryya Gangopadhyay} 
is a Professor in the Department of Information Systems and the Director of the Center for Real-time Distributed Sensing and Autonomy (cards.umbc.edu) at University of Maryland, Baltimore
County. His research interests include Machine Learning and cybersecurity. His research has been funded by NSF, ARL, IBM, and the US Department of Education.
\end{IEEEbiography}

 \vskip -2\baselineskip plus -1fil 
\begin{IEEEbiography}[{\includegraphics[width=1in,height=1.25in,clip,keepaspectratio]{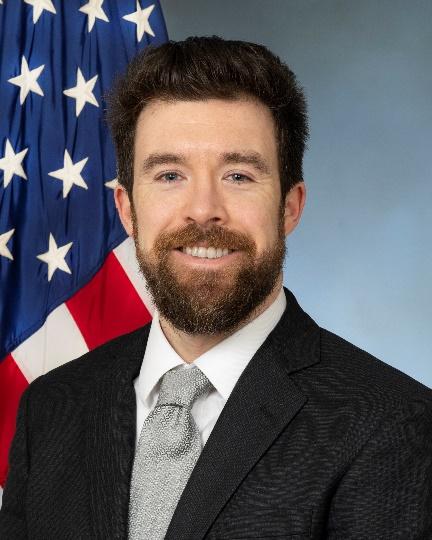}}]{Carl Busart} 
received the B.S. and M.S. degrees from Johns Hopkins University, an MBA from the University of Maryland, College Park, and the D.Eng. degree from George Washington University.  He is a branch chief at the U.S. Army Research Laboratory and a member of the Institute of Electrical and Electronics Engineers (IEEE) and the Association for Computing Machinery (ACM).   His research interests include artificial intelligence / machine learning (AI/ML) and secure design.
\end{IEEEbiography}

 \vskip -2\baselineskip plus -1fil 
\begin{IEEEbiography}[{\includegraphics[width=1in,height=1.25in,clip,keepaspectratio]{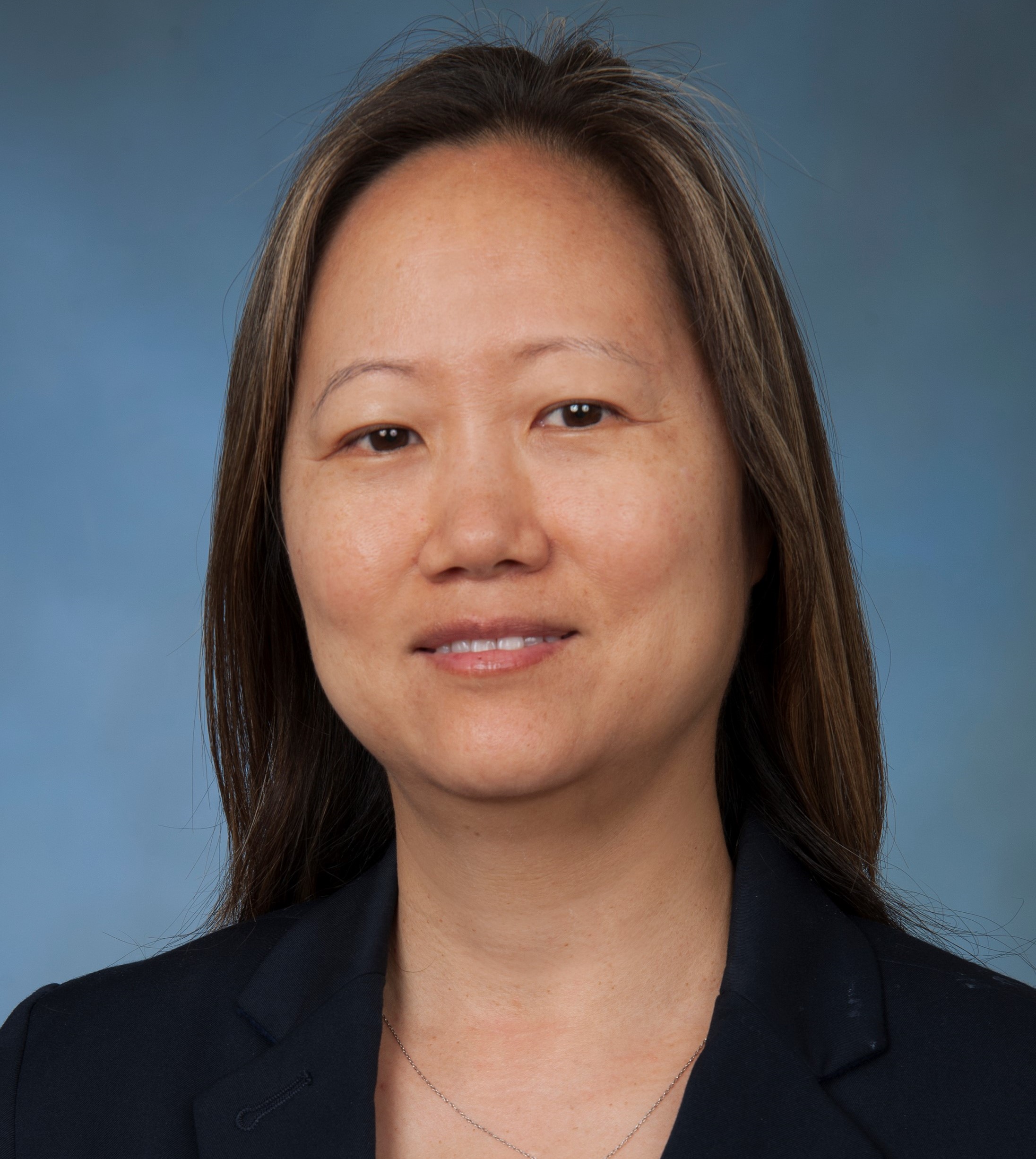}}]{Jade Freeman} 
is the Chief of Battlefield Information Systems Branch at DEVCOM Army Research Laboratory. Her research interest includes information systems for decision support, human-information interactions, and information theory.
\end{IEEEbiography}

 \vskip -2\baselineskip plus -1fil 
\begin{IEEEbiography}[{\includegraphics[width=1in,height=1.25in,clip,keepaspectratio]{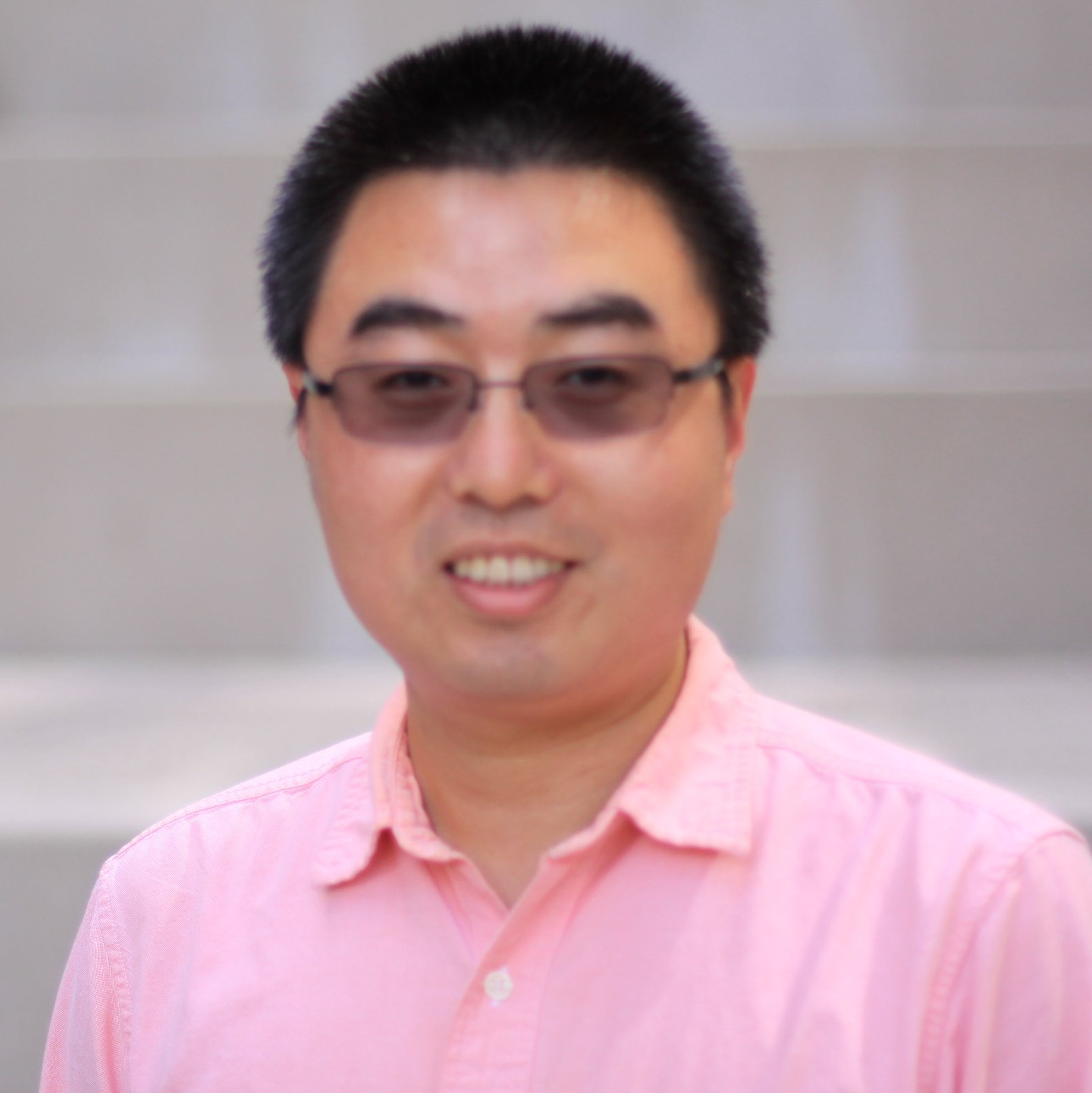}}]{Jianwu Wang}
	 is an Associate Professor of Data Science and the Director of the Big Data Analytics Lab at the Department of Information Systems, University of Maryland, Baltimore County. His research interests include Big Data Analytics, Scientific Workflow, Distributed Computing, Service Oriented Computing. He has published 120+ papers with more than 2700 citations (h-index: 25).
\end{IEEEbiography}



\end{document}

%% file: figure/CR_time.tex
\begin{tikzpicture}[scale=0.6]
\begin{axis}[
	xlabel={Parallelism},
	ylabel={Execution time (h)},
	xtick=data,
	ymin=0,
	ymax=1.2,
	legend columns=2
]
\addplot coordinates {
(	1	,	0.711821431	)
(	2	,	0.461554968	)
(	4	,	0.354325111	)
(	6	,	0.310929292	)
(	8	,	0.293237903	)
};

\addplot [red,mark=square*] coordinates{
(	1	,	0.761618772	)
(	2	,	0.483140932	)
(	4	,	0.362395838	)
(	6	,	0.318328095	)
(	8	,	0.296153406	)
};

\addplot [blue,mark=triangle*] coordinates {
(	1	,	0.899844848	)
(	2	,	0.543417973	)
(	4	,	0.381962341	)
(	6	,	0.335845619	)
(	8	,	0.324151722	)
};

\addplot [red,mark=diamond*] coordinates{
(	1	,	0.869373919	)
(	2	,	0.523031963	)
(	4	,	0.374389532	)
(	6	,	0.321645735	)
(	8	,	0.310015168	)
};

\legend{$\mathsf{AWS~c5d.4xlarge}$, $\mathsf{Azure~F16s\_v2}$, $\mathsf{AWS~c5d.large}$, $\mathsf{Azure~F2s\_v2}$}
\end{axis}
\end{tikzpicture}

%% file: figure/CR_cost.tex
\begin{tikzpicture}[scale=0.6]
\begin{axis}[
	xlabel={Parallelism},
	ylabel={Budgetary cost (\$)},
	xtick=data,
	ymin=0,
	ymax=2.2,
	legend columns=1
]
\addplot [blue,mark=otimes*] coordinates {
(	1	,	0.7297	)
(	2	,	0.5375	)
(	4	,	0.4551	)
(	6	,	0.4218	)
(	8	,	0.4082	)
};


\addplot [dashed,blue,mark=otimes] coordinates {
(	1	,	0.2513348574	)
(	2	,	0.2716185538	)
(	4	,	0.3190608426	)
(	6	,	0.3620952719	)
(	8	,	0.4082067096	)
};

\addplot [red,mark=square*] coordinates {
(	1	,	1.1007	)
(	2	,	0.9041	)
(	4	,	0.8189	)
(	6	,	0.7877	)
(	8	,	0.7721	)
};

\addplot [dashed,red,mark=square] coordinates {
(	1	,	0.6302128566	)
(	2	,	0.6482743745	)
(	4	,	0.6909257309	)
(	6	,	0.7315547263	)
(	8	,	0.7721	)
};

\addplot [blue,mark=triangle*] coordinates {
(	1	,	0.2694	)
(	2	,	0.2973	)
(	4	,	0.3597	)
(	6	,	0.4264	)
(	8	,	0.5019	)
};

\addplot [red,mark=diamond*] coordinates{
(	1	,	0.6618	)
(	2	,	0.6818	)
(	4	,	0.7331	)
(	6	,	0.7822	)
(	8	,	0.8447	)
};

\legend{$\mathsf{AWS~c5d.4xlarge}$, $\mathsf{c5d.4xlarge\_by\_usage}$, $\mathsf{Azure~F16s\_v2}$, $\mathsf{F16s\_v2\_by\_usage}$, $\mathsf{AWS~c5d.large}$, $\mathsf{Azure~F2s\_v2}$}


\end{axis}
\end{tikzpicture}

%% file: figure/CR_ratio.tex
\begin{tikzpicture}[scale=0.6]
\begin{axis}[
	xlabel={Parallelism},
	ylabel={Performance-price ratio},
	xtick=data,
	ymin=0,
	ymax=1.1
]

\addplot [blue,mark=otimes*] coordinates {
(	1	,	0.51940105	)
(	2	,	0.248073894	)
(	4	,	0.161261042	)
(	6	,	0.131148015	)
(	8	,	0.11970168	)
};

\addplot [dashed,blue,mark=otimes] coordinates {
(	1	,	0.178905538	)
(	2	,	0.1253668928	)
(	4	,	0.1130512684	)
(	6	,	0.1125860264	)
(	8	,	0.1197016796	)
};

\addplot [red,mark=square*] coordinates {
(	1	,	0.838315955	)
(	2	,	0.436806508	)
(	4	,	0.296748362	)
(	6	,	0.250759657	)
(	8	,	0.228655396	)
};

\addplot [dashed,red,mark=square] coordinates {
(	1	,	0.4799819418	)
(	2	,	0.3132078855	)
(	4	,	0.2503886093	)
(	6	,	0.2328744224	)
(	8	,	0.2286553961	)
};

\addplot [blue,mark=triangle*] coordinates {
(	1	,	0.242404799	)
(	2	,	0.161577863	)
(	4	,	0.137381747	)
(	6	,	0.143220383	)
(	8	,	0.162707478	)
};

\addplot [red,mark=diamond*] coordinates{
(	1	,	0.575317647	)
(	2	,	0.356620381	)
(	4	,	0.274473429	)
(	6	,	0.251602144	)
(	8	,	0.261882766	)
};

\legend{$\mathsf{AWS~c5d.4xlarge}$, $\mathsf{c5d.4xlarge\_by\_usage}$, $\mathsf{Azure~F16s\_v2}$, $\mathsf{F16s\_v2\_by\_usage}$, $\mathsf{AWS~c5d.large}$, $\mathsf{Azure~F2s\_v2}$}

\end{axis}
\end{tikzpicture}

%% file: figure/CD_time.tex
\begin{tikzpicture}[scale=0.6]
\begin{axis}[
	xlabel={Parallelism},
	ylabel={Execution time (h)},
	xtick=data
]
\addplot [blue,mark=otimes*] coordinates {
(	1	,	0.93611111	)
(	2	,	0.4852777	)
(	4	,	0.2668055	)
(	6	,	0.189722222	)
(	8	,	0.148611111	)

};

\addplot [blue,mark=triangle*] coordinates{
(	1	,	1.0211111	)
(	2	,	0.506944444	)
(	4	,	0.262777778	)
(	6	,	0.186666666	)
(	8	,	0.164722222	)

};

\legend{$\mathsf{AWS~c5d.4xlarge}$,  $\mathsf{AWS~c5d.xlarge}$}
\end{axis}
\end{tikzpicture}

%% file: figure/CD_cost.tex
\begin{tikzpicture}[scale=0.6]
\begin{axis}[
	xlabel={Parallelism},
	ylabel={Budgetary cost (\$)},
	xtick=data,
	ymax=1.1,
]
\addplot [blue,mark=otimes*] coordinates {
(	1	,	0.891933333	)
(	2	,	0.545693274	)
(	4	,	0.377906624	)
(	6	,	0.318706667	)
(	8	,	0.287133333	)
};

\addplot [dashed,blue,mark=otimes] coordinates {
(	1	,	0.2628666666	)
(	2	,	0.2661733184	)
(	4	,	0.275453312	    )
(	6	,	0.2822799999	)
(	8	,	0.2871333333	)
};

\addplot [blue,mark=triangle*] coordinates{
(	1	,	0.3690533312 )
(	2	,	0.367666667	)
(	4	,	0.374813334	)
(	6	,	0.388039999	)
(	8	,	0.426013333	)
};

\legend{$\mathsf{AWS~c5d.4xlarge}$, $\mathsf{c5d.4xlarge\_by\_usage}$,
$\mathsf{AWS~c5d.xlarge}$}
\end{axis}
\end{tikzpicture}

%% file: figure/CD_ratio.tex
\begin{tikzpicture}[scale=0.6]
\begin{axis}[
	xlabel={Parallelism},
	ylabel={Performance-price ratio},
	xtick=data,
	ymax=1.1,
]
\addplot coordinates {
(	1	,	0.834948702	)
(	2	,	0.264812777	)
(	4	,	0.100827566	)
(	6	,	0.060465737	)
(	8	,	0.042671204	)
};

\addplot [dashed,blue,mark=otimes] coordinates {
(	1	,	0.246072407 	)
(	2	,	0.1291679758	)
(	4	,	0.07349245863	)
(	6	,	0.0535547888	)
(	8	,	0.0426712037	)
};

\addplot [blue,mark=triangle*] coordinates{
(	1	,	0.376844453	)
(	2	,	0.186386574	)
(	4	,	0.098492615	)
(	6	,	0.072434133	)
(	8	,	0.070173863	)

};

\legend{ $\mathsf{AWS~c5d.4xlarge}$, $\mathsf{c5d.4xlarge\_by\_usage}$,
$\mathsf{AWS~c5d.xlarge}$}
\end{axis}
\end{tikzpicture}

%% file: figure/DA_time.tex
\begin{tikzpicture}[scale=0.6]
\begin{axis}[
	xlabel={Parallelism},
	ylabel={Execution time (h)},
	xtick=data,
	ymin=0,
	ymax=1.28,
	legend columns=2,
	xtick={1,2,4,6,8},
]
\addplot coordinates {
(	1	,	0.845533611	)
(	2	,	0.667845	)
(	4	,	0.600417778	)

};

\addplot [red,mark=square*] coordinates{
(	1	,	0.731821083	)
(	2	,	0.594235528	)
(	4	,	0.550649694	)

};

\addplot [blue,mark=triangle*] coordinates {
(	1	,	0.84414778	)
(	2	,	0.758208335	)
(	4	,	0.505485558	)
(	6	,	0.398002502	)
(	8	,	0.341232224	)

};

\addplot [red,mark=diamond*] coordinates{
(	1	,	0.97791025	)
(	2	,	0.726843583	)
(	4	,	0.68833775	)
(	6	,	0.548361917	)
(	8	,	0.443764972	)

};

\legend{$\mathsf{AWS\_p3.8xlarge}$, $\mathsf{Azure\_NC24s\_v3}$, $\mathsf{AWS\_p3.2.xlarge}$, $\mathsf{Azure\_NC6s\_v3}$}
\end{axis}
\end{tikzpicture}

%% file: figure/DA_cost.tex
\begin{tikzpicture}[scale=0.6]
\begin{axis}[
	xlabel={Parallelism},
	ylabel={Budgetary cost (\$)},
	xtick=data,
	ymin=0,
	ymax=25,
	legend columns=1,
	xtick={1,2,4,6,8},
]
\addplot [blue,mark=otimes*] coordinates {
(	1	,	10.5323314	)
(	2	,	8.3574228	)
(	4	,	7.5321136	)
};

\addplot [dashed,blue,mark=otimes] coordinates {
(	1	,	2.77033285)
(	2	,	4.2702114)
(	4	,	7.5321136	)
};

\addplot [red,mark=square*] coordinates {
(	1	,	9.52049006	)
(	2	,	7.83644286	)
(	4	,	7.30295226	)
};

\addplot [dashed,red,mark=square] coordinates {
(	1	,	2.802372515	)
(	2	,	4.19972143	)
(	4	,	7.30295226	)
};

\addplot [blue,mark=triangle*] coordinates {
(	1	,	2.766092206	)
(	2	,	4.833235012	)
(	4	,	6.400143223	)
(	6	,	7.540325935	)
(	8	,	8.606364847	)
};

\addplot [red,mark=diamond*] coordinates{
(	1	,	3.555405365	)
(	2	,	5.01128273	)
(	4	,	8.98825406	)
(	6	,	10.63092479	)
(	8	,	11.42636652	)
};

\legend{$\mathsf{AWS\_p3.8xlarge}$, $\mathsf{p3.8xlarge\_by\_usage}$, $\mathsf{Azure\_NC24s\_v3}$, $\mathsf{NC24s\_v3\_by\_usage}$,  $\mathsf{AWS\_p3.2.xlarge}$, $\mathsf{Azure\_NC6s\_v3}$}


\end{axis}
\end{tikzpicture}

%% file: figure/DA_ratio.tex
\begin{tikzpicture}[scale=0.6]
\begin{axis}[
	xlabel={Parallelism},
	ylabel={Performance-price ratio},
	xtick=data,
	ymin=0,
	ymax=15,
	xtick={1,2,4,6,8},
]

\addplot [blue,mark=otimes*] coordinates {
(	1	,	8.905440202	)
(	2	,	5.58146303	)
(	4	,	4.52241491	)
};

\addplot [dashed,blue,mark=otimes] coordinates {
(	1	,	2.342409539)
(	2	,   2.851839332)
(	4	,	4.52241491)
};

\addplot [red,mark=square*] coordinates {
(	1	,	6.96729535	)
(	2	,	4.656692759	)
(	4	,	4.021368431	)
};

\addplot [dashed,red,mark=square] coordinates {
(	1	,	2.05083529	)
(	2	,	2.49562368	)
(	4	,	4.021368431)
};

\addplot [blue,mark=triangle*] coordinates {
(	1	,	2.334990594	)
(	2	,	3.664599072	)
(	4	,	3.235179965	)
(	6	,	3.001068587	)
(	8	,	2.936769018	)
};

\addplot [red,mark=diamond*] coordinates{
(	1	,	3.476867349	)
(	2	,	3.642418697	)
(	4	,	6.186954576	)
(	6	,	5.829594294	)
(	8	,	5.070621221	)
};

\legend{$\mathsf{AWS_p3.8xlarge}$, $\mathsf{p3.8xlarge\_by\_usage}$, $\mathsf{Azure\_NC24s\_v3}$, $\mathsf{NC24s\_v3\_by\_usage}$,  $\mathsf{AWS_p3.2.xlarge}$, $\mathsf{Azure\_NC6s\_v3}$}

\end{axis}
\end{tikzpicture}

%% file: figure/SC_time.tex
\begin{tikzpicture}[scale=0.6]
\begin{axis}[
	xlabel={Parallelism},
	ylabel={Execution time (h)},
	xtick=data
]
\addplot [blue,mark=otimes*] coordinates {
(	1	,	23.67	)
(	2	,	12.84	)
(	4	,	8.27	)
(	6	,	6.33	)
(	8	,	6.01	)

};

\addplot [blue,mark=triangle*] coordinates{
(	1	,	25.6	)
(	2	,	14.14	)
(	4	,	9.08	)
(	6	,	7.18	)
(	8	,	6.61	)

};

\legend{$\mathsf{AWS\_c5d.4xlarge}$,  $\mathsf{AWS\_c5d.xlarge}$}
\end{axis}
\end{tikzpicture}

%% file: figure/SC_cost.tex
\begin{tikzpicture}[scale=0.6]
\begin{axis}[
	xlabel={Parallelism},
	ylabel={Budgetary cost (\$)},
	xtick=data,
	ymax=20,
]
\addplot [blue,mark=otimes*] coordinates {
(	1	,	19.36	)
(	2	,	11.04	)
(	4	,	7.54	)
(	6	,	6.05	)
(	8	,	5.79	)
};

\addplot [dashed,blue,mark=otimes] coordinates {
(	1	,	3.45	)
(	2	,	3.65	)
(	4	,	4.36    )
(	6	,	4.83	)
(	8	,	5.79	)
};

\addplot [blue,mark=triangle*] coordinates{
(	1	,	 3.64   )
(	2	,	5.08	)
(	4	,	8.12	)
(	6	,	11.23	)
(	8	,	14.54	)
};

\legend{$\mathsf{AWS\_c5d.4xlarge}$, $\mathsf{c5d.4xlarge\_by\_usage}$,
$\mathsf{AWS\_c5d.xlarge}$}
\end{axis}
\end{tikzpicture}

%% file: figure/SC_ratio.tex
\begin{tikzpicture}[scale=0.6]
\begin{axis}[
	xlabel={Parallelism},
	ylabel={Performance-price ratio},
	xtick=data,
	ymax=500,
]
\addplot coordinates {
(	1	,	458.14	)
(	2	,	141.73	)
(	4	,	62.36	)
(	6	,	38.28	)
(	8	,	34.77	)
};

\addplot [dashed,blue,mark=otimes] coordinates {
(	1	,	81.77 	)
(	2	,	46.82	)
(	4	,	36.07	)
(	6	,	30.58	)
(	8	,	34.77	)
};

\addplot [blue,mark=triangle*] coordinates{
(	1	,	93.21	)
(	2	,	71.83	)
(	4	,	74.57	)
(	6	,	80.69	)
(	8	,	96.20	)

};

\legend{ $\mathsf{AWS\_c5d.4xlarge}$, $\mathsf{c5d.4xlarge\_by\_usage}$,
$\mathsf{AWS\_c5d.xlarge}$}
\end{axis}
\end{tikzpicture}

%% file: figure/CR_boxplot.tex








\begin{tikzpicture}[scale=0.47]
  \begin{axis}
    [
    ylabel={Execution time (h)},
    boxplot/draw direction=y,
    ytick distance=0.05,
    xtick={1,2,3,4,5,6},
    xticklabels={CR\\w/o Rep., CR\\w Rep., CD\\w/o Rep., CD\\w Rep., DA\\w/o Rep., DA\\w Rep.},
    x=1.2cm,
    ticklabel style = {font=\footnotesize},
    label style={font=\small},
    cycle list={{red},{blue}},
    x tick label style={font=\small, text width=3cm, align=center}
    ]
    \addplot+[color=red,
    boxplot prepared={
      median=0.34055,
      average=0.34129256352857,
      upper quartile=0.34944,
      lower quartile=0.3319444444,
      upper whisker=0.35432,
      lower whisker=0.3301824663
    },
    ] coordinates {};
    \addplot+[color=blue,
    boxplot prepared={
      median=0.3489848121,
      average=0.34771319845714,
      upper quartile=0.35666,
      lower quartile=0.3405555556,
      upper whisker=0.3607,
      lower whisker=0.3352777778
    },
    ] coordinates {};
    \addplot+[color=red,
    boxplot prepared={
      median=0.265,
      average=0.2689768425,
      upper quartile=0.2815,
      lower quartile=0.26124,
      upper whisker=0.2912,
      lower whisker=0.26
    },
    ] coordinates {};
    \addplot+[color=blue,
    boxplot prepared={
      median=0.275,
      average=0.2789768425,
      upper quartile=0.28929166625,
      lower quartile=0.265104125,
      upper whisker=0.299,
      lower whisker=0.26
    },
    ] coordinates {};
    \addplot+[color=red,
    boxplot prepared={
      median=0.5971,
      average=0.6281155721,
      upper quartile=0.66788719275,
      lower quartile=0.59440719375,
      upper whisker=0.7486641454,
      lower whisker=0.5822332974
    },
    ] coordinates {};
    \addplot+[color=blue,
    boxplot prepared={
      median=0.60946166835,
      average=0.63817112765,
      upper quartile=0.682151123025,
      lower quartile=0.5972504507,
      upper whisker=0.7383863676,
      lower whisker=0.5854260722
    },
    ] coordinates {};
  \end{axis}
\end{tikzpicture}

%% file: figure/CR_boxplot_relative.tex
\begin{tikzpicture}[scale=0.45]
\begin{axis}[
    every axis plot post/.style={/pgf/number format/fixed},
    ybar,
    bar width=16pt,
    x=2.3cm,
    ymin=0,
    axis on top,
    ymin=0.5,
    ymax=1.2,
    xtick=data,
    enlarge x limits=0.2,
    ticklabel style = {font=\small},
    label style={font=\small},
    symbolic x coords={CR,CD,DA},
    visualization depends on=rawy\as\rawy, 
    after end axis/.code={ 
        \draw [ultra thick, white, decoration={snake, amplitude=1pt}, decorate] (rel axis cs:0,1.0) -- (rel axis cs:1,1.0);},
    nodes near coords={
        \pgfmathprintnumber{\rawy}},
    every node near coord/.append style={font=\footnotesize},
    axis lines*=left,
    clip=false,
    legend columns=2,
    legend image code/.code={\draw[#1] (0cm,-0.1cm) rectangle (0.5cm,0.1cm);},
    legend style={at={(0.05,1.1)},
    anchor=north west},
]
 
\addplot
coordinates 
{(CR,0.981534682) 
(CD,0.964154731)
(DA,0.984243168)};
\addplot [draw=blue,pattern=horizontal lines light blue] coordinates {(CR,1)(CD,1)(DA,1)};

\legend{$\textit{w/o~Rep.}$,$\textit{w~Rep.}$}

\end{axis}
\end{tikzpicture}

%% file: figure/SDK_boxplot.tex
\begin{tikzpicture}[scale=0.47]
  \begin{axis}
    [
    ylabel={Execution time (h)},
    boxplot/draw direction=y,
    ytick distance=0.05,
    xtick={1,2,3,4,5,6},
    xticklabels={CR w\\SDK, CR w\\Serverless, CD w\\SDK, CD w\\Serverless, DA w\\SDK, DA w\\Serverless},
    x=1.2cm,
    ticklabel style = {font=\footnotesize},
    label style={font=\small},
    cycle list={{red},{blue}},
    x tick label style={font=\small, text width=3cm, align=center}
    ]
    \addplot+[color=red,
    boxplot prepared={
      median=0.43455178281,
      average=0.43782945309,
      upper quartile=0.445109769025,
      lower quartile=0.4290690015,
      upper whisker=0.45286765000,
      lower whisker=0.42581739100
    },
    ] coordinates {};
    \addplot+[color=blue,
    boxplot prepared={
      median=0.3489848121,
      average=0.34771319845714,
      upper quartile=0.35666,
      lower quartile=0.3405555556,
      upper whisker=0.3607,
      lower whisker=0.3352777778
    },
    ] coordinates {};
    \addplot+[color=red,
    boxplot prepared={
      median=0.36249556375,
      average=0.35777818,
      upper quartile=0.3695094675,
      lower quartile=0.351665625,
      upper whisker=0.374687,
      lower whisker=0.34576978
    },
    ] coordinates {};
    \addplot+[color=blue,
    boxplot prepared={
      median=0.275,
      average=0.2789768425,
      upper quartile=0.28929166625,
      lower quartile=0.265104125,
      upper whisker=0.299,
      lower whisker=0.26
    },
    ] coordinates {};
    \addplot+[color=red,
    boxplot prepared={
      median=0.83754362,
      average=0.82583127665556,
      upper quartile=0.84854454725,
      lower quartile=0.80301143265,
      upper whisker=0.8676547625,
      lower whisker=0.7976235476
    },
    ] coordinates {};
    \addplot+[color=blue,
    boxplot prepared={
      median=0.60946166835,
      average=0.63817112765,
      upper quartile=0.682151123025,
      lower quartile=0.5972504507,
      upper whisker=0.7383863676,
      lower whisker=0.5854260722
    },
    ] coordinates {};
  \end{axis}
\end{tikzpicture}

%% file: figure/SDK_boxplot_relative.tex
\begin{tikzpicture}[scale=0.45]
\begin{axis}[
    every axis plot post/.style={/pgf/number format/fixed},
    ybar,
    bar width=16pt,
    x=2.3cm,
    ymin=0,
    axis on top,
    ymax=1.5,
    xtick=data,
    enlarge x limits=0.2,
    ticklabel style = {font=\small},
    label style={font=\small},
    symbolic x coords={CR,CD,DA},
    visualization depends on=rawy\as\rawy, 
    after end axis/.code={ 
        \draw [ultra thick, white, decoration={snake, amplitude=1pt}, decorate] (rel axis cs:0,1.0) -- (rel axis cs:1,1.0);},
    nodes near coords={
        \pgfmathprintnumber{\rawy}},
    every node near coord/.append style={font=\footnotesize},
    axis lines*=left,
    clip=false,
    legend columns=2,
    legend image code/.code={\draw[#1] (0cm,-0.1cm) rectangle (0.5cm,0.1cm);},
    legend style={at={(0.05,1.1)},
    anchor=north west},
]
 
\addplot coordinates {(CR,1.259168346)(CD,1.282465515)(DA,1.29405929)};
\addplot [draw=blue,pattern=horizontal lines light blue] coordinates {(CR,1)(CD,1)(DA,1)};

\legend{$\textit{w~SDK}$,$\textit{w~Serverless}$}

\end{axis}
\end{tikzpicture}